\documentclass{IEEEtran}
\usepackage{cite}
\usepackage{amsmath,amssymb,amsfonts}
\usepackage{algorithmic}
\usepackage{graphicx}
\usepackage{soul}
\usepackage{xcolor}
\usepackage{textcomp}
\usepackage{lipsum}

\ifCLASSOPTIONcompsoc
    \usepackage[caption=false, font=normalsize, labelfont=sf, textfont=sf]{subfig}
\else
    \usepackage[caption=false, font=footnotesize]{subfig}
\fi

\def\BibTeX{{\rm B\kern-.05em{\sc i\kern-.025em b}\kern-.08em
    T\kern-.1667em\lower.7ex\hbox{E}\kern-.125emX}}
\begin{document}
\title{Field Synthesis with Azimuthally-Varying, Cascaded, Cylindrical Metasurfaces using a Wave Matrix Approach}

\author{Chun-Wen Lin, \IEEEmembership{Student Member, IEEE}, and Anthony Grbic, \IEEEmembership{Fellow, IEEE}

\thanks{This work has been submitted to the IEEE for possible publication. Copyright may be transferred without notice, after which this version may no longer be accessible.}

\thanks{This work was supported by the Office of Naval Research (ONR) under Grant N00014-18-1-2536, the Air Force Office of Scientific Research (AFOSR) Multidisciplinary University Research Initiative (MURI) program under Grant FA9550-18-1-0379, and the Chia-Lun Lo Fellowship at the University of Michigan. \textit{(Corresponding Author: Anthony Grbic.)}}

\thanks{The authors are with the Radiation Laboratory, Department of Electrical Engineering and Computer Science, University of Michigan, Ann Arbor, MI 48109-2122 USA (email: chunwen@umich.edu; agrbic@umich.edu)}
}

\maketitle


\begin{abstract}
In recent years, there has been extensive research on planar metasurfaces capable of arbitrarily controlling scattered fields. However, rigorous studies on conformal metasurfaces, such as those that are cylindrical, have been few in number likely due to their more complex geometry and corresponding analysis. Here, wave propagation in cascaded cylindrical structures consisting of layers of dielectric spacers and azimuthally-varying metasurfaces (subwavelength patterned metallic claddings) is investigated. A wave matrix approach, which incorporates the advantages of both ABCD (transmission) matrices and scattering matrices ($S$ matrices), is adopted. Wave matrices are used to model the higher-order coupling between metasurface layers, overcoming the fabrication difficulties associated with previous works. The proposed framework provides an efficient approach to synthesize the inhomogeneous sheet admittances that realize desired cylindrical field transformations. Design examples are reported to illustrate the power and potential applications of the proposed method in antenna design and stealth technology.

\end{abstract}



\section{Introduction}
\label{sec:introduction}
\IEEEPARstart{C}{ylindrical} metasurfaces, a popular category of conformal metasurfaces, have been widely utilized in scenarios where planar metasurfaces are not applicable due to mechanical, aerodynamic or hydrodynamic reasons. Their ability to tailor both the amplitude and phase of cylindrical waves finds use in radiation pattern control [\ref{Raeker_2015}-\ref{Raeker_2016}], scattering control [\ref{Sipus_2018}-\ref{Vellucci_2017}], cloaking [\ref{Soric_2015}-\ref{Xu_2021}], illusion [\ref{Xu_2021}-\ref{Kwon_2020}], high gain antenna design [\ref{Xu_2021}], beam steering [\ref{Li_2019}] and angular momentum generation [\ref{Li(2)_2019}]. 

Many electromagnetic design problems involve transforming a known excitation field to a desired radiation field. In field transformations, conversion between azimuthal modes is often necessary, and bianisotropic responses are readily needed including electric, magnetic, and magneto-electric effects [\ref{Xu_2020}]. In order to achieve azimuthal mode conversion, cylindrical metasurfaces are modeled with bianisotropic polarizability tensors in [\ref{Safari_2019}]. The synthesis of a field transformation simply involves finding the required surface polarization current densities that match the tangential components of the source field to the desired field. Nevertheless, complex polarizabilities, which lead to negative resistances, are unavoidable with this technique. However, by utilizing surface waves in design, local power conservation can be satisfied [\ref{Epstein_2016}], and field transformation realized with passive and lossless cylindrical metasurfaces [\ref{Kwon_2020}]. Still, the reported method in [\ref{Kwon_2020}] mandates that metasurfaces be impenetrable and of zero thickness simultaneously. Such metasurfaces become unrealizable in practice. Due to the challenges associated with realizing magnetic and magneto-electric effects, bianisotropic responses are typically implemented by cascading layers of metasurfaces [\ref{Pfeiffer_2013}-\ref{Pfeiffer_2014}]. In [\ref{Xu_2021}, \ref{Li(2)_2019}-\ref{Xu_2020}], bianisotropic responses required for field transformations are first derived by a mode-matching technique, and then later replaced by cascaded structures. Although neat and elegant, directly replacing idealized bianisotropic boundaries with cascaded structures leads to two significant issues for realization. First, to minimize the effects due to finite thickness in cascaded structures, the separation between metasurfaces has to be extremely small [\ref{Xu_2021}, \ref{Xu_2020}]. These small separation distances minimize transverse propagation and coupling within the cascaded metasurfaces. Alternatively, perfect conducting baffles need to be inserted to prevent higher order azimuthal modes from propagating between layers [\ref{Xu_2021}, \ref{Li(2)_2019}-\ref{Xu_2020}]. Both of these requirements greatly increase fabrication complexity as well as cost.

\begin{figure}[!t]
\centerline{\includegraphics[width=8.4cm]{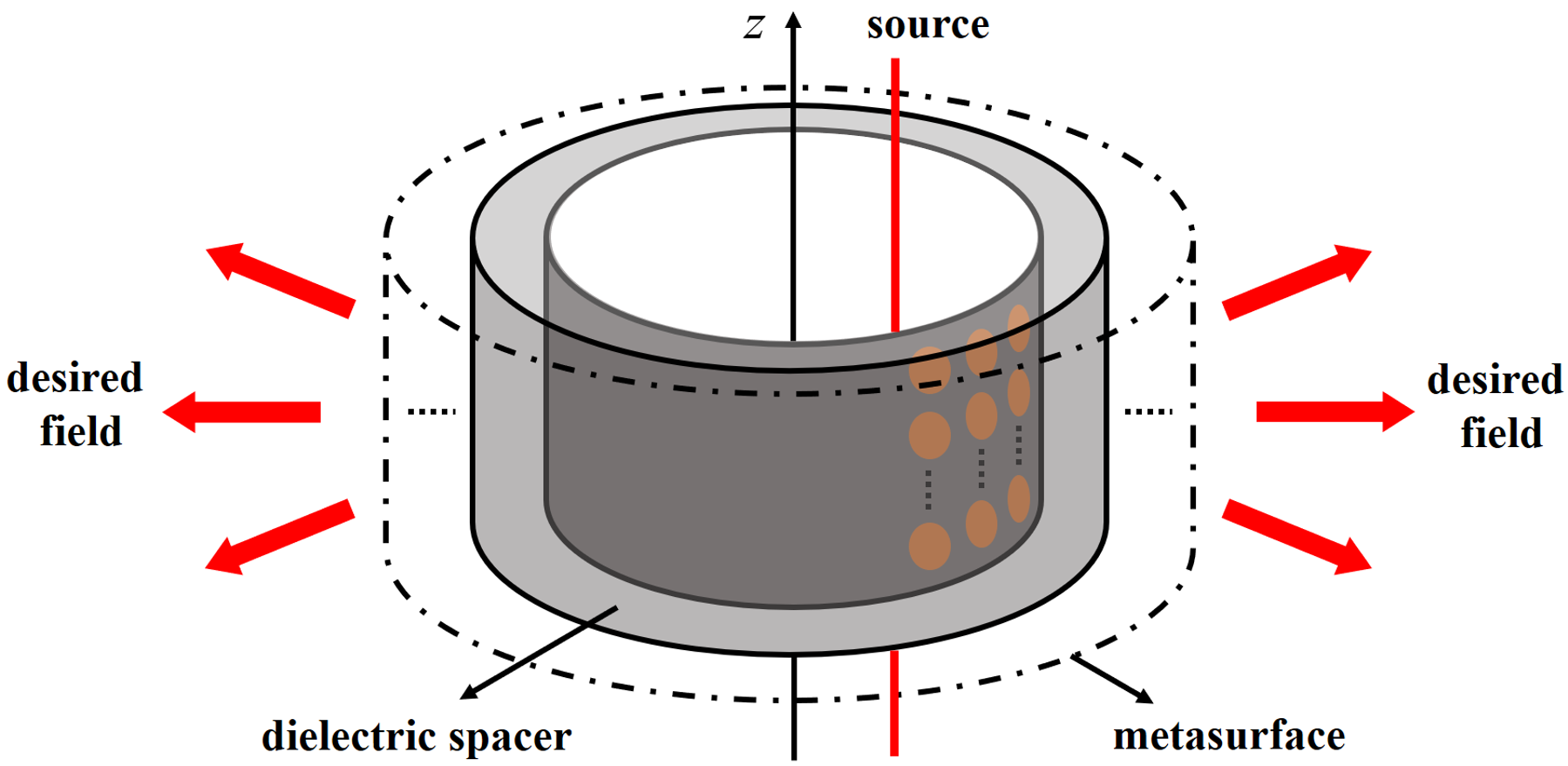}}
\caption{Illustration of the cascaded, concentric cylindrical metasurfaces [\ref{Sipus_2018}, \ref{my_TAP}, \ref{my_APS}]. Wave propagation along the radial direction is studied.}
\label{fig1}
\end{figure}

A more rigorous approach to analyzing cascaded cylindrical metasurfaces is to start from network parameters, where multiple azimuthal modes can be accounted for simultaneously. Hence, higher-order coupling between layers can be intrinsically modeled, and the aforementioned fabrication difficulties circumvented. An ABCD (transmission) matrix formulation in cylindrical coordinates can be adopted to accurately capture the radial wave propagation in cascaded structures [\ref{Sipus_2018}]. However, ABCD matrices involve total electric and magnetic fields and thus are not well suited for synthesizing the scattering properties of the metasurfaces. A wave matrix approach, relating the incident and scattered electric fields on one side of the metasurfaces to those on the other side [\ref{Collin}], is a better choice.
Not only do wave matrices directly provide scattered field information, but they also allow for the simple analysis of cascaded structures through matrix multiplication. In [\ref{Ranjbar_2017}], this method was exploited to design cascaded, planar metasurfaces with arbitrarily specified $S$ matrices. The wave matrix theory for azimuthally invariant structures in cylindrical coordinates has been derived in [\ref{my_TAP}] to model the cascaded metasurfaces illustrated in Fig. \ref{fig1}. The power of this method was demonstrated by synthesizing interesting azimuthally invariant devices such as polarization converter and polarization splitter analytically.

As noted, only azimuthally-invariant metasurfaces, which do not involve conversion between azimuthal modes, were investigated in this previous work [\ref{my_TAP}]. Since fields along a cylindrical geometry can be decomposed into elementary cylindrical waves with different azimuthal orders, arbitrary field transformation is not possible with merely azimuthally-invariant structures. In order to address this issue, we combine the concepts of multimodal matrix analysis [\ref{Alsolamy_2021}] and the mode matching technique [\ref{Sipus_2019}]. By considering multiple azimuthal modes simultaneously, the proposed theory can be generalized, and azimuthal mode converters successfully designed [\ref{my_APS}]. 

In this paper, the mathematical background for multimodal wave-matrix theory [\ref{my_APS}] is discussed in detail. We aim to accomplish arbitrary field transformations through the design of azimuthally-varying, cascaded, cylindrical metasurfaces. Azimuthal variations introduce interactions and conversions between different azimuthal modes, which are absent in the case of single-mode, azimuthally-invariant metasurfaces [\ref{my_TAP}]. First, the definition of multimodal wave matrices, ABCD matrices, $S$ matrices in cylindrical coordinates, along with the conversion formulas between them are provided. Moreover, the wave matrix expressions of the building blocks that make up the cascaded structures are derived. Based on the proposed theory, an optimization process is employed to determine the metasurface admittances required for a stipulated field transformation. As verification, several design examples, including azimuthal mode converters, illusion devices, and multi-functional metasurfaces are illustrated and verified through commercial electromagnetic solvers. In particular, multi-input-multi-output functionality has not been demonstrated before in cylindrical metasurface structures. Lastly, suggestions on selecting reasonable design parameters, as well as guidelines on practical realization are provided.

The proposed devices do not assume idealized bianisotropic boundary conditions nor extremely close metasurface separations (which are challenging to fabricate and can lead to higher-order coupling between metasurfaces). In addition, the devices do not require conducting baffles to isolate unit cells. The design examples showcase the capability of the proposed approach in various applications.




\section{The Wave Matrix Theory}
\label{sec:theory}

In this section, the basic assumptions and formulation employed in this paper to derive the wave matrix theory are introduced. The definition of network parameters and conversion formulas between them [\ref{my_TAP}] are generalized to account for multiple azimuthal modes.

\subsection{Problem Setup}
\begin{figure}[!t]
\centerline{\includegraphics[width=6.5cm]{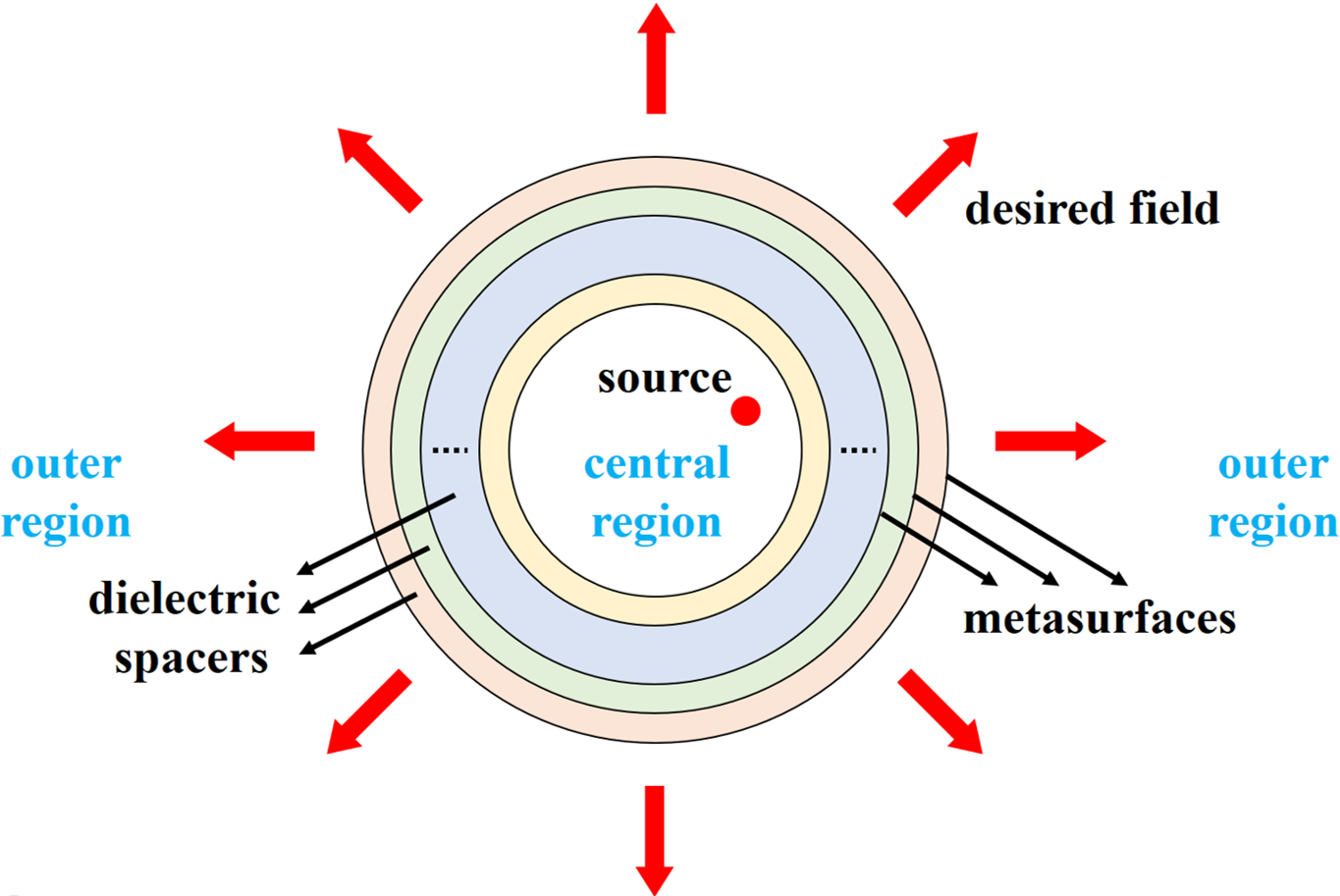}}
\caption{The 2-D problem of concern, which is a cross-section of Fig. \ref{fig1}. Without loss of generality, it is assumed that the sources are located in the central region of the cascaded metasurfaces.}
\label{fig2}
\end{figure}

Throughout this paper, a two-dimensional scenario is assumed in order to simplify the problems considered. All structures, including the excitation, metasurfaces, and dielectric spacers are independent of $z$, and there is no propagation constant along the $z$ direction ($k_z=0$). Therefore, a cross-section of Fig. \ref{fig1} along the $z=0$ plane, shown in Fig. \ref{fig2}, is sufficient to describe all electromagnetic field distributions. Furthermore, as shown in Fig. \ref{fig2}, it is also assumed that the cascaded structure is excited by electric currents, resulting in TM$_z$ waves. The case of TE$_z$ waves can be derived directly from duality.

As in [\ref{my_TAP}, \ref{my_APS}], the synthesis and analysis of the metasurfaces are conducted in the spectral domain. The time convention of $e^{+j\omega t}$ is assumed, and is suppressed in all subsequent expressions. For TM$_z$ waves with $k_z=0$, the only field components tangential to the metasurfaces are $E_z$ as well as $H_\phi$. The electric field $E_z$ can be written as a summation over all azimuthal modes:
\begin{equation}
\begin{split}
    E_z(\rho,\phi) &= \sum_{m=-M}^{+M} 
    \mathcal{E}_m(\rho) e^{-jm\phi}
    \\
    &= \sum_{m=-M}^{+M} \mathcal{E}_m^+(\rho) e^{-jm\phi} 
    + \sum_{m=-M}^{+M} \mathcal{E}_m^-(\rho) e^{-jm\phi}
.\end{split}
\label{Ez_def}\end{equation}

\noindent{In the above equation, $M$ is a sufficiently large number that ensures the series converges. Generally speaking, as we have more layers of metasurfaces, or as the metasurfaces have more complicated spatial variations, $M$ also needs to be larger. The quantity $\mathcal{E}_m$ is the $m^{th}$ azimuthal mode of the total electric field, while $\mathcal{E}_m^+$ and $\mathcal{E}_m^-$ are its outward and inward propagating parts, respectively. Since $H_m^{(2)}$ and $H_m^{(1)}$ represent outward and inward propagating waves respectively under $e^{+j\omega t}$ convention, we can write:
}
\begin{equation}
\begin{split}
    \mathcal{E}_m^+(\rho) &= \alpha_m^+ H_m^{(2)}(k\rho) \\
    \mathcal{E}_m^-(\rho) &= \alpha_m^- H_m^{(1)}(k\rho)
.\end{split}
\label{Ez_propagating}\end{equation}

\noindent{where $\alpha_m^+$ and $\alpha_m^-$ represent outward and inward traveling wave amplitudes for azimuthal order $m$. These wave amplitudes are generally different in each dielectric layer. Similarly, the tangential magnetic field can be derived [\ref{Harrington}]:}
\begin{equation}
    H_\phi(\rho,\phi) =  \sum_{m=-M}^{+M} 
    \mathcal{H}_m(\rho) e^{-jm\phi}
\label{Hp_def}\end{equation}
\begin{equation}
    \mathcal{H}_m(\rho) = 
    -\frac{j\omega\varepsilon}{k}
    \Big{[}\alpha_m^+ H_m^{(2)'}(k\rho) + \alpha_m^- H_m^{(1)'}(k\rho)\Big{]}
.\label{Hp_total}\end{equation}

\noindent{These azimuthal mode quantities will be used in the definition of the network parameters, as discussed in the next part.}

\subsection{Definition of Network Parameters}
\begin{figure}[!t]
\centerline{\includegraphics[width=6.5cm]{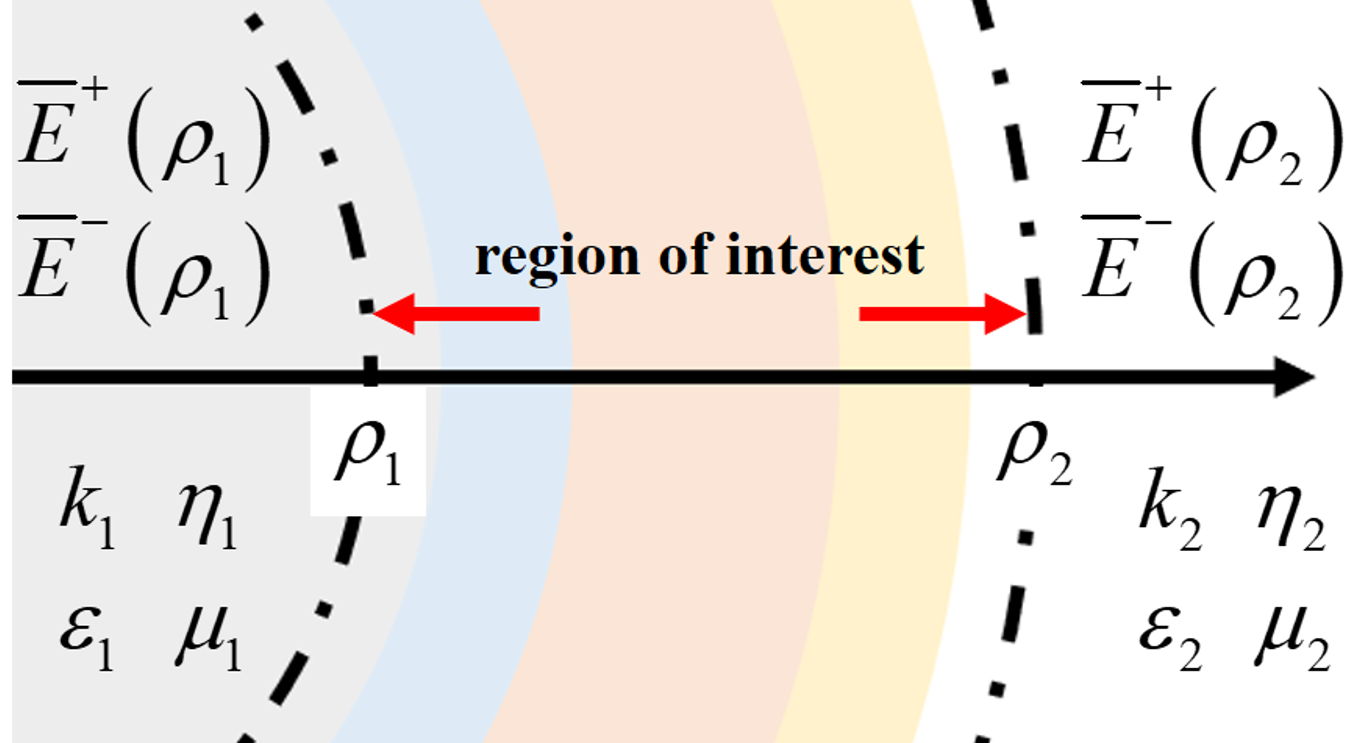}}
\caption{The annulus region enclosed by two cylindrical surfaces (denoted by their radii $\rho_1$ and $\rho_2$) is the region of interest. Network parameters relate the azimuthal modes of the fields at these two radii.}
\label{fig3}
\end{figure}

Network parameters, commonly used to analyze and characterize microwave circuits [\ref{Collin}, \ref{Pozar}], have also proven very useful for solving field problems. Consider the region of interest in Fig. \ref{fig3} enclosed by cylindrical surfaces with an inner radius (inner port) $\rho_1$ and an outer radius (outer port) $\rho_2$. A wave matrix is defined to relate the incident and reflected waves on radius $\rho_1$ to those at radius $\rho_2$,

\begin{equation}
    \begin{bmatrix}
    \bar{E}^+(\rho_1) \\ \bar{E}^-(\rho_1)
    \end{bmatrix}
    =
    \begin{bmatrix}
    \bar{\bar{W}}_{++} & \bar{\bar{W}}_{+-} \\ \bar{\bar{W}}_{-+} & \bar{\bar{W}}_{--}
    \end{bmatrix}
    \cdot
    \begin{bmatrix}
    \bar{E}^+(\rho_2) \\ \bar{E}^-(\rho_2)
    \end{bmatrix}
\label{eq_Wave_Matrix_def}\end{equation}

\noindent{where the column vectors $\bar{E}^\pm(\rho)$ are composed of azimuthal modes of the propagating electric fields at radius $\rho$:}
\begin{equation}
    \begin{split}
        \bar{E}^+(\rho) &= 
        [\mathcal{E}_M^+(\rho), ..., \mathcal{E}_0^+(\rho), ..., \mathcal{E}_{-M}^+(\rho)]^T
        \\
        \bar{E}^-(\rho) &= 
        [\mathcal{E}_M^-(\rho), ..., \mathcal{E}_0^-(\rho), ..., \mathcal{E}_{-M}^-(\rho)]^T
    .\end{split}
\label{eq_Wave_Vector_def}\end{equation}

\noindent{Since $\bar{E}^\pm(\rho)$ are both $(2M+1)\times 1$ vectors, the matrices $\bar{\bar{W}}_{++}$, $\bar{\bar{W}}_{+-}$, $\bar{\bar{W}}_{-+}$, and $\bar{\bar{W}}_{--}$ are of dimension $(2M+1)\times(2M+1)$. Accordingly, the wave matrix is a $(4M+2)\times(4M+2)$ matrix. In [\ref{my_TAP}], only one azimuthal mode is considered, so the wave matrix in (\ref{eq_Wave_Matrix_def}) reduces to a $2\times 2$ matrix for TM$_z$ waves.}

Likewise, we define an ABCD matrix by relating the azimuthal modes of total electric and magnetic fields on the two cylindrical surfaces:
\begin{equation}
    \begin{bmatrix}
    \bar{E}(\rho_1) \\ \bar{H}(\rho_1)
    \end{bmatrix}
    =
    \begin{bmatrix}
    \bar{\bar{A}} & \bar{\bar{B}} \\ \bar{\bar{C}} & \bar{\bar{D}}
    \end{bmatrix}
    \cdot
    \begin{bmatrix}
    \bar{E}(\rho_2) \\ \bar{H}(\rho_2)
    \end{bmatrix}
\label{eq_ABCD_Matrix_def}\end{equation}

\noindent{where the column vectors}
\begin{equation}
    \begin{split}
        \bar{E}(\rho) &= 
        [\mathcal{E}_M(\rho), ..., \mathcal{E}_0(\rho), ..., \mathcal{E}_{-M}(\rho)]^T
        \\
        \bar{H}(\rho) &= 
        [\mathcal{H}_M(\rho), ..., \mathcal{H}_0(\rho), ..., \mathcal{H}_{-M}(\rho)]^T
    .\end{split}
\label{eq_ABCD_Vector_def}\end{equation}

\noindent{As in the case of a wave matrix, the dimension of the ABCD matrix is also $(4M+2)\times(4M+2)$. The total field used to define the ABCD matrices are directly related to the boundary conditions. Hence, these matrices will be extensively studied when complicated metasurface boundaries are involved.}

On the other hand, a scattering matrix ($S$ matrix) in cylindrical coordinates bears a more complex form. For the cases of planar metasurfaces, an $S$ matrix is just a rearrangement of its wave matrix counterpart, since the wave impedance is identical everywhere. In cylindrical coordinates, the wave impedance depends not only on the radius [\ref{my_TAP}], but also the azimuthal mode of concern [\ref{Harrington}]. Normalization is required to ensure that the $S$ matrix is unitary when the system is lossless, and symmetric when the system is reciprocal. A reasonable definition of an $S$ matrix is based on the $\rho$-directed TM$_z$ power [\ref{my_TAP}, \ref{Kurokawa_1965}]. By generalizing the single mode $S$ matrix derivation and expression in [\ref{my_TAP}], we obtain:
\begin{equation}
    \begin{bmatrix}
    \bar{\bar{c}}_1^- \bar{E}^-(\rho_1) \\ 
    \bar{\bar{c}}_2^+ \bar{E}^+(\rho_2)
    \end{bmatrix}
    =
    \begin{bmatrix}
    \bar{\bar{S}}_{11} & \bar{\bar{S}}_{12} \\
    \bar{\bar{S}}_{21} & \bar{\bar{S}}_{22}
    \end{bmatrix}
    \cdot
    \begin{bmatrix}
    \bar{\bar{c}}_1^+ \bar{E}^+(\rho_1) \\
    \bar{\bar{c}}_2^- \bar{E}^-(\rho_2)
    \end{bmatrix}
.\label{eq_S_Matrix_def}\end{equation}

\noindent{The matrices $\bar{\bar{c}}_1^\pm$ and $\bar{\bar{c}}_2^\pm$ are introduced for compactness and algebraic purposes. For example, the diagonal matrix $\bar{\bar{c}}_i^+$ contains the normalization information for outward propagating waves at radius $\rho_i$:}
\begin{equation}
    \bar{\bar{c}}_i^+ =
    \begin{bmatrix}
    c_{i,M}^+ & & \hdots & & 0 \\
    & \ddots & & & \\
    \vdots & & c_{i,0}^+ & & \vdots \\
    & & & \ddots & \\
    0 & & \hdots & & c_{i,-M}^+
  \end{bmatrix}
\label{eq_c_matrix_def}\end{equation}

\noindent{in which the power normalization coefficients $c_{i,m}^+$ is related to the radius $\rho_i$ and the azimuthal mode $m$ of concern [\ref{my_TAP}]}
\begin{equation}
    c_{i,m}^+ = \sqrt{2\pi\rho_i \text{Re}
    \bigg\{
    \frac{j\omega\varepsilon_i}{k_i}\frac{H_m^{(2)'}(k_i\rho_i)}{H_m^{(2)}(k_i\rho_i)}
    \bigg\}} 
.\label{eq_c_coeff_def}\end{equation}

\noindent{Finally, the diagonal matrix $\bar{\bar{c}}_i^-$ is similarly defined for inward propagating waves at $\rho_i$:}
\begin{equation}
    \bar{\bar{c}}_i^- =
    \begin{bmatrix}
    c_{i,M}^- & & \hdots & & 0 \\
    & \ddots & & & \\
    \vdots & & c_{i,0}^- & & \vdots \\
    & & & \ddots & \\
    0 & & \hdots & & c_{i,-M}^-
  \end{bmatrix}
\label{eq_c_matrix_def_2}\end{equation}
\begin{equation}
    c_{i,m}^- = \sqrt{2\pi\rho_i \text{Re}
    \bigg\{
    \frac{j\omega\varepsilon_i}{k_i}\frac{H_m^{(1)'}(k_i\rho_i)}{H_m^{(1)}(k_i\rho_i)}
    \bigg\}} 
.\label{eq_c_coeff_def_2}\end{equation}


\subsection{Conversion Formulas between Network Parameters}
After defining the network parameters in cylindrical coordinates, conversion formulas between them can also be derived through algebraic manipulation. To convert from $S$ matrices to wave matrices, we multiply out the first row of the block matrix equation (\ref{eq_S_Matrix_def}),
\begin{equation}
    \bar{\bar{c}}_1^- \bar{E}^-(\rho_1)
    = \bar{\bar{S}}_{11} \bar{\bar{c}}_1^+ \bar{E}^+(\rho_1) 
    + \bar{\bar{S}}_{12} \bar{\bar{c}}_2^- \bar{E}^-(\rho_2)
.\label{eq_S_wave_derive_1}\end{equation}
\noindent{This equation (\ref{eq_S_wave_derive_1}) can be rearranged to the following}
\begin{equation}
    -\bar{\bar{S}}_{11} \bar{\bar{c}}_1^+ \bar{E}^+(\rho_1) 
    + \bar{\bar{c}}_1^- \bar{E}^-(\rho_1) 
    = \bar{\bar{O}} \bar{E}^+(\rho_2)
    + \bar{\bar{S}}_{12} \bar{\bar{c}}_2^- \bar{E}^-(\rho_2)
.\label{eq_S_wave_derive_2}\end{equation}
\noindent{where $\bar{\bar{O}}$ represents a zero matrix of dimension $(2M+1)\times(2M+1)$. Analogously, the second row of (\ref{eq_S_Matrix_def}) becomes}
\begin{equation}
    -\bar{\bar{S}}_{21} \bar{\bar{c}}_1^+ \bar{E}^+(\rho_1) 
    + \bar{\bar{O}} \bar{E}^-(\rho_1) 
    = -\bar{\bar{c}}_2^+ \bar{E}^+(\rho_2)
    + \bar{\bar{S}}_{22} \bar{\bar{c}}_2^- \bar{E}^-(\rho_2)
.\label{eq_S_wave_derive_3}\end{equation}
\noindent{Combining (\ref{eq_S_wave_derive_2}) and (\ref{eq_S_wave_derive_3}) yields a matrix equation}
\begin{equation}
    \begin{bmatrix}
        -\bar{\bar{S}}_{11} \bar{\bar{c}}_1^+ & \bar{\bar{c}}_1^- \\
        -\bar{\bar{S}}_{21} \bar{\bar{c}}_1^+ & \bar{\bar{O}}
    \end{bmatrix}
    \begin{bmatrix}
        \bar{E}^+(\rho_1) \\ \bar{E}^-(\rho_1)
    \end{bmatrix}
    =
    \begin{bmatrix}
        \bar{\bar{O}} & \bar{\bar{S}}_{12} \bar{\bar{c}}_2^- \\
        -\bar{\bar{c}}_2^+ & \bar{\bar{S}}_{22} \bar{\bar{c}}_2^-
    \end{bmatrix}
    \begin{bmatrix}
        \bar{E}^+(\rho_2) \\ \bar{E}^-(\rho_2)
    \end{bmatrix}
.\label{eq_S_wave_derive_4}\end{equation}
\noindent{By comparing (\ref{eq_S_wave_derive_4}) with the definition of a wave matrix (\ref{eq_Wave_Matrix_def}), we arrive at the conversion formula from $S$ matrices to wave matrices as}
\begin{equation}
\begin{bmatrix}
    \bar{\bar{W}}_{++} & \bar{\bar{W}}_{+-} \\ \bar{\bar{W}}_{-+} & \bar{\bar{W}}_{--}
    \end{bmatrix}
    =
    \begin{bmatrix}
        -\bar{\bar{S}}_{11} \bar{\bar{c}}_1^+ & \bar{\bar{c}}_1^- \\
        -\bar{\bar{S}}_{21} \bar{\bar{c}}_1^+ & \bar{\bar{O}}
    \end{bmatrix}^{-1}
    \begin{bmatrix}
        \bar{\bar{O}} & \bar{\bar{S}}_{12} \bar{\bar{c}}_2^- \\
        -\bar{\bar{c}}_2^+ & \bar{\bar{S}}_{22} \bar{\bar{c}}_2^-
    \end{bmatrix}
.\label{S2Wave}\end{equation}

One can also express an $S$ matrix in terms of a wave matrix by expanding (\ref{eq_Wave_Matrix_def}). For instance, the first row of this block matrix equation is equivalent to
\begin{equation}
    \bar{\bar{O}} \bar{E}^-(\rho_1) - \bar{\bar{W}}_{++} \bar{E}^+(\rho_2) =
    -\bar{\bar{I}} \bar{E}^+(\rho_1) + \bar{\bar{W}}_{+-} \bar{E}^-(\rho_2)
\label{eq_wave_S_derive_1}\end{equation}

\noindent{where $\bar{\bar{I}}$ stands for an $(2M+1)\times(2M+1)$ identity matrix. In order to align (\ref{eq_wave_S_derive_1}) with the definition of $S$ matrices (\ref{eq_S_Matrix_def}), we rewrite the above equation as}
\begin{equation}
\begin{split}
    &\bar{\bar{O}} (\bar{\bar{c}}_1^-)^{-1} \bar{\bar{c}}_1^- \bar{E}^-(\rho_1) 
    - \bar{\bar{W}}_{++} (\bar{\bar{c}}_2^+)^{-1} \bar{\bar{c}}_2^+ \bar{E}^+(\rho_2)
    \\ &=
    -\bar{\bar{I}} (\bar{\bar{c}}_1^+)^{-1} \bar{\bar{c}}_1^+ \bar{E}^+(\rho_1) 
    + \bar{\bar{W}}_{+-} (\bar{\bar{c}}_2^-)^{-1} \bar{\bar{c}}_2^- \bar{E}^-(\rho_2)
\end{split}
.\label{eq_wave_S_derive_2}\end{equation}

\noindent{Following the same procedure by which (\ref{S2Wave}) is derived, we obtain the following conversion formula from wave matrices to $S$ matrices,}
\begin{equation}
\begin{split}
    &\begin{bmatrix}
        \bar{\bar{S}}_{11} & \bar{\bar{S}}_{12} \\
        \bar{\bar{S}}_{21} & \bar{\bar{S}}_{22}
    \end{bmatrix}
    \\ &=
    \begin{bmatrix}
        \bar{\bar{O}} & -\bar{\bar{W}}_{++} (\bar{\bar{c}}_2^+)^{-1} \\
        (\bar{\bar{c}}_1^-)^{-1} & -\bar{\bar{W}}_{-+} (\bar{\bar{c}}_2^+)^{-1}
    \end{bmatrix}^{-1}
    \begin{bmatrix}
        -(\bar{\bar{c}}_1^+)^{-1} & \bar{\bar{W}}_{+-} (\bar{\bar{c}}_2^-)^{-1} \\
        \bar{\bar{O}} & \bar{\bar{W}}_{--} (\bar{\bar{c}}_2^-)^{-1}
    \end{bmatrix}
.\end{split}
\label{Wave2S}\end{equation}

On the other hand, the relation between wave matrices and ABCD matrices can be easily obtained by generalizing the result in [\ref{my_TAP}]. In [\ref{my_TAP}], a transformation matrix has been defined that relates the incident and reflected electric fields to the total electric and magnetic fields on a cylindrical surface (denoted by a single radius $\rho$). For TM$_z$ waves, we have the single azimuthal mode transformation matrix [\ref{my_TAP}]
\begin{equation}
    \begin{bmatrix}
        \mathcal{E}_m(\rho) \\ \mathcal{H}_m(\rho)
    \end{bmatrix}
    =
    \begin{bmatrix}
        1 & 1 \\
        -Y_m^+ & Y_m ^-
    \end{bmatrix}
    \cdot
    \begin{bmatrix}
        \mathcal{E}^+_m(\rho) \\ \mathcal{E}^-_m(\rho)
    \end{bmatrix}
\label{eq_transformation_single}\end{equation}
\begin{equation}
Y_m^+ = +\frac{j}{\eta} \frac{H_m^{(2)'}(k\rho)}{H_m^{(2)}(k\rho)} ,\quad
Y_m^- = -\frac{j}{\eta} \frac{H_m^{(1)'}(k\rho)}{H_m^{(1)}(k\rho)}
.\label{eq_T_single}\end{equation}

\noindent{In fact, $Y_m^+$ is exactly the outward wave admittance, while $Y_m^-$ is the inward wave admittance of azimuthal order $m$ at radius $\rho$ [\ref{Harrington}]. For the case of multiple modes, the network parameters are defined by superposing each azimuthal mode. As a result, the multimodal transformation matrix can be expressed as:}
\begin{equation}
    \begin{bmatrix}
    T_{(k,\eta)}(\rho)
    \end{bmatrix}
    = 
    \begin{bmatrix}
        \bar{\bar{I}} & \bar{\bar{I}} \\
        -\bar{\bar{Y}}^+ & \bar{\bar{Y}}^-
    \end{bmatrix}
\label{eq_transformation_multimodal}\end{equation}

\noindent{where the $(2M+1)\times(2M+1)$ diagonal matrices $\bar{\bar{Y}}^+$ and $\bar{\bar{Y}}^-$ consist of entries $Y_m^+$ or $Y_m^-$ for each mode}
\begin{equation}
    \bar{\bar{Y}}^+ 
    =
    \begin{bmatrix}
    Y_M^+ & \hdots & 0 \\
    \vdots & \ddots & \vdots \\
    0 & \hdots & Y_{-M}^+
    \end{bmatrix}
    ,\text{ }
    \bar{\bar{Y}}^- 
    =
    \begin{bmatrix}
    Y_M^- & \hdots & 0 \\
    \vdots & \ddots & \vdots \\
    0 & \hdots & Y_{-M}^-
    \end{bmatrix}
.\label{eq_T_multimodal}\end{equation}

\noindent{The transformation matrix (\ref{eq_transformation_multimodal}) satisfies the following relation}
\begin{equation}
    \begin{bmatrix}
        \bar{E}(\rho) \\ \bar{H}(\rho)
    \end{bmatrix}
    =
    \begin{bmatrix}
    T_{(k,\eta)}(\rho)
    \end{bmatrix}
    \cdot
    \begin{bmatrix}
        \bar{E}^+(\rho) \\ \bar{E}^-(\rho)
    \end{bmatrix}
.\label{eq_transformation_relation}\end{equation}

\noindent{Therefore, the definition of an ABCD matrix (\ref{eq_ABCD_Matrix_def}) can be rewritten by incorporating (\ref{eq_transformation_relation}),}
\begin{equation}
\begin{split}
    &\begin{bmatrix}
        T_{(k_1,\eta_1)}(\rho_1)
    \end{bmatrix}
    \begin{bmatrix}
        \bar{E}^+(\rho_1) \\ \bar{E}^-(\rho_1)
    \end{bmatrix}
    \\ &=
    \begin{bmatrix}
        \bar{\bar{A}} & \bar{\bar{B}} \\ \bar{\bar{C}} & \bar{\bar{D}}
    \end{bmatrix}
    \begin{bmatrix}
        T_{(k_2,\eta_2)}(\rho_2)
    \end{bmatrix}
    \begin{bmatrix}
        \bar{E}^+(\rho_2) \\ \bar{E}^-(\rho_2)
    \end{bmatrix}
.\end{split}
\end{equation}

\noindent{The definition of a wave matrix (\ref{eq_Wave_Matrix_def}) directly implies the conversion formula from ABCD matrices to wave matrices}
\begin{equation}
    \begin{bmatrix}
        \bar{\bar{W}}_{++} & \bar{\bar{W}}_{+-} \\ \bar{\bar{W}}_{-+} & \bar{\bar{W}}_{--}
    \end{bmatrix}
    =
    \begin{bmatrix}
        T_{(k_1,\eta_1)}(\rho_1)
    \end{bmatrix}^{-1}
    \begin{bmatrix}
        \bar{\bar{A}} & \bar{\bar{B}} \\ \bar{\bar{C}} & \bar{\bar{D}}
    \end{bmatrix}
    \begin{bmatrix}
        T_{(k_2,\eta_2)}(\rho_2)
    \end{bmatrix}
,\label{ABCD2Wave}\end{equation}

\noindent{as well as the one from wave matrices to ABCD matrices}
\begin{equation}
    \begin{bmatrix}
        \bar{\bar{A}} & \bar{\bar{B}} \\ \bar{\bar{C}} & \bar{\bar{D}}
    \end{bmatrix}
    =
    \begin{bmatrix}
        T_{(k_1,\eta_1)}(\rho_1)
    \end{bmatrix}
    \begin{bmatrix}
        \bar{\bar{W}}_{++} & \bar{\bar{W}}_{+-} \\ \bar{\bar{W}}_{-+} & \bar{\bar{W}}_{--}
    \end{bmatrix}
    \begin{bmatrix}
        T_{(k_2,\eta_2)}(\rho_2)
    \end{bmatrix}^{-1}
.\label{Wave2ABCD}\end{equation}


\section{Building Blocks of the Cascaded Cylindrical Structure}
\label{sec:blocks}
Based on the aforementioned theory, the wave matrices describing the electromagnetic properties of dielectric spacers, dielectric interfaces and azimuthally-varying metasurfaces are determined in this section. The wave matrix of the cascaded system, which may be very complicated if derived by brute-force, can be easily obtained from the wave matrices of building blocks by multiplying them in sequential order.

\subsection{Dielectric Spacers}
\begin{figure}[!t]
\centerline{\includegraphics[width=5cm]{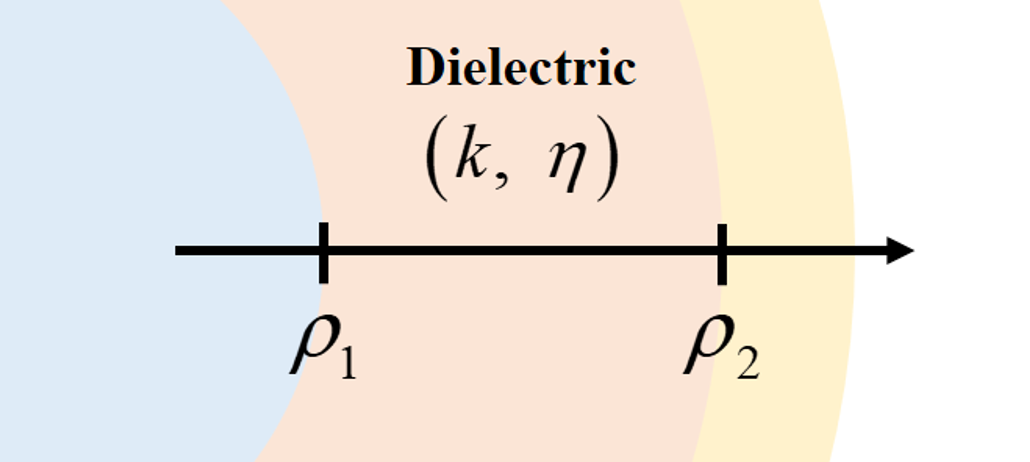}}
\caption{A dielectric spacer with properties $(k,\eta)$. The innermost radius of the dielectric layer is $\rho_1$, while the outermost radius is denoted as $\rho_2$.}
\label{fig4}
\end{figure}

Fig. \ref{fig4} illustrates a dielectric spacer bounded by two cylindrical surfaces with radii $\rho_1$ and $\rho_2$. We assume that the dielectric is isotropic as well as homogeneous. Accordingly, no azimuthal mode mixing occurs. Each mode propagates independently. This implies that every wave amplitude $\alpha_m^\pm$ remains the same throughout the dielectric spacer. We can relate $\mathcal{E}_m^\pm(\rho_1)$ and $\mathcal{E}_m^\pm(\rho_2)$ by considering (\ref{Ez_propagating}):
\begin{equation}
    \mathcal{E}_m^+(\rho_1) = F_m^+ \cdot \mathcal{E}_m^+(\rho_2) 
    ,\quad
    \mathcal{E}_m^-(\rho_1) = F_m^- \cdot \mathcal{E}_m^-(\rho_2)
\end{equation}
\noindent{with the quantities}
\begin{equation}
    F_m^+ = \frac{ H_m^{(2)}(k\rho_1) }{ H_m^{(2)}(k\rho_2) }
    ,\quad
    F_m^- = \frac{ H_m^{(1)}(k\rho_1) }{ H_m^{(1)}(k\rho_2) }
.\label{eq_H_def}\end{equation}

\noindent{By juxtaposing each azimuthal mode, the wave matrix of a dielectric spacer can be written in the form of}
\begin{equation}
    \begin{bmatrix}
        W
    \end{bmatrix}_D
    =
    \begin{bmatrix}
        \bar{\bar{W}}_{++} & \bar{\bar{W}}_{+-}
        \\
        \bar{\bar{W}}_{-+} & \bar{\bar{W}}_{--}
    \end{bmatrix}
    =
    \begin{bmatrix}
        \bar{\bar{F}}^+ & \bar{\bar{O}}
        \\
        \bar{\bar{O}} & \bar{\bar{F}}^-
    \end{bmatrix}
,\label{Wave_Dielectric}\end{equation}

\noindent{where the $(2M+1)\times(2M+1)$ diagonal matrices $\bar{\bar{F}}^+$ and $\bar{\bar{F}}^-$ are defined as}
\begin{equation}
    \bar{\bar{F}}^+ 
    =
    \begin{bmatrix}
    F_M^+ & \hdots & 0 \\
    \vdots & \ddots & \vdots \\
    0 & \hdots & F_{-M}^+
    \end{bmatrix}
    ,\text{ }
    \bar{\bar{F}}^- 
    =
    \begin{bmatrix}
    F_M^- & \hdots & 0 \\
    \vdots & \ddots & \vdots \\
    0 & \hdots & F_{-M}^-
    \end{bmatrix}
.\end{equation}

It is worth mentioning that in the limiting case where the radii are very large, the Hankel functions behave similarly to the exponential functions. The entries of the wave matrix (\ref{eq_H_def}) become just phase delays between the input and output ports, which is identical to the case of a planar dielectric slab [\ref{Ranjbar_2017}].

\subsection{Dielectric Interfaces}
\begin{figure}[!t]
\centerline{\includegraphics[width=4.5cm]{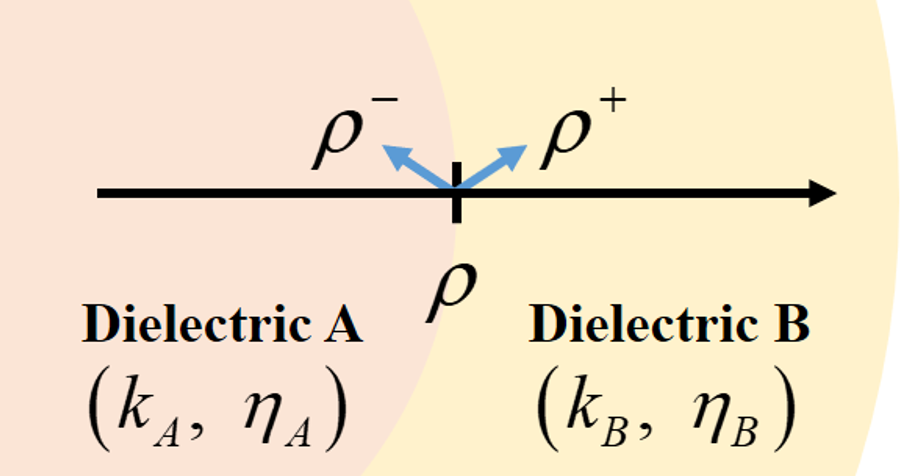}}
\caption{Example of an interface. The interface can either be a simple dielectric interface or one with a metasurface.}
\label{fig5}
\end{figure}

Consider the interface (a cylindrical surface with radius $\rho$) between two dielectric materials as shown in Fig. \ref{fig5}. The value $\rho^-$ is taken to be infinitely close to but smaller than $\rho$, so that it falls in the inner dielectric. Contrarily, the value $\rho^+$ is defined to be infinitely close to but larger than $\rho$, so that it belongs to the outer dielectric. Since no surface current density exists on the interface, there is continuity of the tangential electric and magnetic fields,
\begin{equation}
    E_z(\rho^-,\phi) = E_z(\rho^+,\phi)
    , \quad
    H_\phi(\rho^-,\phi) = H_\phi(\rho^+,\phi)
.\label{eq_BC_interface}\end{equation}

\noindent{These field quantities in (\ref{eq_BC_interface}) can be written as summations over azimuthal modes, as in (\ref{Ez_def}) and (\ref{Hp_def}). Mode matching implies:}
\begin{equation}
    \mathcal{E}_m(\rho^-) = \mathcal{E}_m(\rho^+)
    , \quad
    \mathcal{H}_m(\rho^-) = \mathcal{H}_m(\rho^+)
\label{eq_mode_matching}\end{equation}

\noindent{for all azimuthal orders $m$. Hence, in vector form (\ref{eq_ABCD_Vector_def}), we write}
\begin{equation}
    \begin{bmatrix}
        \bar{E}(\rho^-) \\ \bar{H}(\rho^-)
    \end{bmatrix}
    =
    \begin{bmatrix}
        \bar{E}(\rho^+) \\ \bar{H}(\rho^+)
    \end{bmatrix}
.\label{eq_BC_interface_mode}\end{equation}

\noindent{Applying transformation matrices (\ref{eq_transformation_multimodal}) to (\ref{eq_BC_interface_mode}), we arrive at}
\begin{equation}
    \begin{bmatrix}
    T_{(k_A,\eta_A)}(\rho^-)
    \end{bmatrix}
    \begin{bmatrix}
        \bar{E}^+(\rho^-) \\ \bar{E}^-(\rho^-)
    \end{bmatrix}
    =
    \begin{bmatrix}
    T_{(k_B,\eta_B)}(\rho^+)
    \end{bmatrix}
    \begin{bmatrix}
        \bar{E}^+(\rho^+) \\ \bar{E}^-(\rho^+)
    \end{bmatrix}
\end{equation}

\noindent{which results in a compact form for the wave matrix of a dielectric interface:}
\begin{equation}
    \begin{bmatrix}
        W
    \end{bmatrix}_I
    =
    \begin{bmatrix}
    T_{(k_A,\eta_A)}(\rho^-)
    \end{bmatrix}^{-1}
    \cdot
    \begin{bmatrix}
    T_{(k_B,\eta_B)}(\rho^+)
    \end{bmatrix}
.\label{Wave_Interface}\end{equation}

\subsection{Metasurfaces}
Now, let us assume that an azimuthally-varying metasurface exists at the cylindrical boundary shown in Fig. \ref{fig5}. In this paper, metasurfaces with only electric responses are considered since these responses can be easily realized by metallic patterning. Mathematically, an electric response is often characterized by an electric admittance which is the ratio of surface current density to averaged electric field [\ref{Tretyakov_2003}]. Since the metasurface is periodic in $\phi$ (with period $2\pi$), we can express its admittance profile $Y_{MS}(\phi)$ as a Fourier series,
\begin{equation}
\begin{split}
    Y_{MS}(\phi) &= p_0 + \sum_{k=1}^K q_k\cos(k\phi) + \sum_{k=1}^K r_k\sin(k\phi)
    \\
    &= p_0 
    + \sum_{k=1}^K (\frac{q_k}{2} + \frac{r_k}{2j}) e^{+jk\phi} 
    + \sum_{k=1}^K (\frac{q_k}{2} - \frac{r_k}{2j}) e^{-jk\phi}
.\end{split}
\label{eq_Y_profile}\end{equation}

\noindent{In (\ref{eq_Y_profile}), the highest azimuthal order $K$, along with the parameters $p_0$, $q_k$, and $r_k$ define the metasurface. There is a trade-off between higher $K$ values and the computational complexity. Larger $K$ values of a single-layer metasurface generate more complicated azimuthal mode mixing. However, more computational resources are required to design the metasurface admittance, as well as the metallic patterning.}

The tangential fields across a metasurface at a radius $\rho$ obey the following boundary conditions
\begin{equation}
E_z(\rho^-,\phi) = E_z(\rho^+,\phi) 
\label{eq_BC_E_metasurface}\end{equation}
\begin{equation}
H_\phi(\rho^-,\phi) = H_\phi(\rho^+,\phi) - Y_{MS}(\phi)E_z(\rho^+,\phi)
.\label{eq_BC_H_metasurface}\end{equation}

\noindent{In order to derive the wave matrix of this metasurface, we start with its ABCD matrix representation,}
\begin{equation}
    \begin{bmatrix}
        \bar{\bar{A}} & \bar{\bar{B}} \\
        \bar{\bar{C}} & \bar{\bar{D}}
    \end{bmatrix}_{MS}
    =
    \begin{bmatrix}
        \bar{\bar{A}}_{MS} & \bar{\bar{B}}_{MS} \\
        \bar{\bar{C}}_{MS} & \bar{\bar{D}}_{MS}
    \end{bmatrix}
\label{eq_ABCD_for_metasurface}\end{equation}
\noindent{which is defined based on the matrix equations (\ref{eq_ABCD_Matrix_def}),}
\begin{equation}
    \bar{E}(\rho^-) = \bar{\bar{A}}_{MS} \bar{E}(\rho^+) + \bar{\bar{B}}_{MS} \bar{H}(\rho^+)
\label{eq_trouble_0(a)}\end{equation}
\begin{equation}
    \bar{H}(\rho^-) = \bar{\bar{C}}_{MS} \bar{E}(\rho^+) + \bar{\bar{D}}_{MS} \bar{H}(\rho^+)
.\label{eq_trouble_0(b)}\end{equation}

\noindent{Our target is to decompose the boundary conditions (\ref{eq_BC_E_metasurface}) and (\ref{eq_BC_H_metasurface}) into their azimuthal modes $\mathcal{E}_m(\rho^\pm)$ and $\mathcal{H}_m(\rho^\pm)$. Next, we cast the azimuthal modes into vector form $\bar{E}(\rho^\pm)$, $\bar{H}(\rho^\pm)$ as defined by (\ref{eq_ABCD_Vector_def}), and substitute these vectors into the matrix equations (\ref{eq_trouble_0(a)}), (\ref{eq_trouble_0(b)}) so that $\bar{\bar{A}}_{MS}$, $\bar{\bar{B}}_{MS}$, $\bar{\bar{C}}_{MS}$ and $\bar{\bar{D}}_{MS}$ can be found.}


First, the electric field boundary condition (\ref{eq_BC_E_metasurface}) can be decomposed as
\begin{equation}
    \sum_m \mathcal{E}_m(\rho^-) e^{-jm\phi} = \sum_m \mathcal{E}_m(\rho^+) e^{-jm\phi}
,\end{equation}

\noindent{indicating that $\mathcal{E}_m(\rho^-) = \mathcal{E}_m(\rho^+)$ for all azimuthal orders $m$. By casting each mode into vector form (\ref{eq_ABCD_Vector_def}), we arrive at}
\begin{equation}
    \bar{E}(\rho^-) = \bar{E}(\rho^+) = \bar{\bar{I}}\bar{E}(\rho^+) + \bar{\bar{O}}\bar{H}(\rho^+)
,\end{equation}

\noindent{which combines with the definition (\ref{eq_trouble_0(a)}) to yield}
\begin{equation}
    \bar{\bar{A}}_{MS} = \bar{\bar{I}}
    , \quad
    \bar{\bar{B}}_{MS} = \bar{\bar{O}}
.\end{equation}
\noindent{Decomposing the fields in (\ref{eq_BC_H_metasurface}) into azimuthal modes gives us}
\begin{equation}
\begin{split}
    &\sum_m \mathcal{H}_m(\rho^-) e^{-jm\phi}
    \\ &=
    \sum_m \mathcal{H}_m(\rho^+) e^{-jm\phi}
    -
    Y_{MS}(\phi) \sum_{m'} \mathcal{E}_{m'}(\rho^+) e^{-jm'\phi}
.\end{split}
\label{eq_trouble_1}\end{equation}

\noindent{The first summation on the right hand side of (\ref{eq_trouble_1}) together with (\ref{eq_trouble_0(b)}) implies}
\begin{equation}
    \bar{\bar{D}}_{MS} = \bar{\bar{I}}
.\end{equation}

\noindent{On the contrary, the expression for $\bar{\bar{C}}_{MS}$ is more complicated since the admittance profile $Y_{MS}(\phi)$ also contains a $\phi$ dependence. Substituting (\ref{eq_Y_profile}) into the second summation on the right hand side of (\ref{eq_trouble_1}) yields}
\begin{equation}
\begin{split}
    &-\sum_{m'} p_0 \cdot \mathcal{E}_{m'}(\rho^+) e^{-jm'\phi} \\
    &-\sum_{m'} \sum_k (\frac{q_k}{2} + \frac{r_k}{2j}) \cdot \mathcal{E}_{m'}(\rho^+) e^{-j(m'-k)\phi} \\
    &-\sum_{m'} \sum_k (\frac{q_k}{2} - \frac{r_k}{2j}) \cdot \mathcal{E}_{m'}(\rho^+) e^{-j(m'+k)\phi}
.\end{split}
\label{eq_long_1}\end{equation}

\noindent{To perform mode matching, the summation variables are changed. In (\ref{eq_long_1}), we let $m'=m$ in the first summation, $m'=m+k$ in the second summation, and $m'=m-k$ in the last one. Accordingly, (\ref{eq_long_1}) becomes}
\begin{equation}
\begin{split}
    &-\sum_m p_0 \cdot \mathcal{E}_m(\rho^+) e^{-jm\phi} \\
    &-\sum_m \sum_k (\frac{q_k}{2} + \frac{r_k}{2j}) \cdot \mathcal{E}_{m+k}(\rho^+) e^{-jm\phi} \\
    &-\sum_m \sum_k (\frac{q_k}{2} - \frac{r_k}{2j}) \cdot \mathcal{E}_{m-k}(\rho^+) e^{-jm\phi}
.\end{split}
\label{eq_long_2}\end{equation}

In matrix form, the first summation in (\ref{eq_long_2}) can be represented by an $(2M+1)\times(2M+1)$ identity matrix $\bar{\bar{I}}$. To express the second summation in (\ref{eq_long_2}) in matrix form, we define a set of new matrices $\bar{\bar{L}}^{(k)}$, all with the same dimension $(2M+1)\times(2M+1)$. The $(i,j)$-th element of $\bar{\bar{L}}^{(k)}$ matrix satisfies
\begin{equation}
    \bar{\bar{L}}^{(k)}(i,j)
    =
    \begin{cases}
        1,  & \text{if }  i>k, \text{and }i=j+k.\\
        0,  & \text{otherwise.}
    \end{cases}
\end{equation}
\noindent{The matrix $\bar{\bar{L}}^{(k)}$ can be viewed as shifting all the ones in an identity matrix $\bar{\bar{I}}$ to the left by $k$ positions. For example, if $M=1$, $2M+1=3$, then}
\begin{equation*}
    \bar{\bar{L}}^{(1)} 
    = 
    \begin{bmatrix}
        0 & 0 & 0 \\
        1 & 0 & 0 \\
        0 & 1 & 0
    \end{bmatrix}
    ,\text{ }
    \bar{\bar{L}}^{(2)} 
    = 
    \begin{bmatrix}
        0 & 0 & 0 \\
        0 & 0 & 0 \\
        1 & 0 & 0
    \end{bmatrix}
.\end{equation*}

\noindent{Similarly, to express the third summation in (\ref{eq_long_2}) in matrix form, another set of $(2M+1)\times(2M+1)$ matrices $\bar{\bar{R}}^{(k)}$ are defined by shifting all the ones in an identity matrix to the right by $k$ positions:}
\begin{equation}
    \bar{\bar{R}}^{(k)}(i,j)
    =
    \begin{cases}
        1,  & \text{if }  i\leq(2M+1)-k, \text{and }i=j-k.\\
        0,  & \text{otherwise.}
    \end{cases}
\end{equation}

\noindent{Truncating the summation range of (\ref{eq_long_2}) from $m=-M$ to $m=+M$ and applying the $\bar{\bar{L}}^{(k)}$ and $\bar{\bar{R}}^{(k)}$ matrices, (\ref{eq_long_2}) can be cast into a matrix form as required in (\ref{eq_trouble_0(b)}):}
\begin{equation}
    \bigg[
    (-p_0)\bar{\bar{I}} 
    - \sum_{k=1}^K (\frac{q_k}{2} + \frac{r_k}{2j}) \bar{\bar{L}}^{(k)}
    - \sum_{k=1}^K (\frac{q_k}{2} - \frac{r_k}{2j}) \bar{\bar{R}}^{(k)}
    \Bigg]
    \bar{E}(\rho^+)
,\end{equation}
\noindent{which indicates that}
\begin{equation}
    \bar{\bar{C}}_{MS} = 
    (-p_0)\bar{\bar{I}} 
    - \sum_{k=1}^K (\frac{q_k}{2} + \frac{r_k}{2j}) \bar{\bar{L}}^{(k)}
    - \sum_{k=1}^K (\frac{q_k}{2} - \frac{r_k}{2j}) \bar{\bar{R}}^{(k)}
.\end{equation}

\noindent{Finally, the wave matrix of a metasurface described by an admittance profile (\ref{eq_Y_profile}) can be derived by converting from the ABCD matrix using (\ref{ABCD2Wave}),}
\begin{equation}
    \begin{bmatrix}
        W
    \end{bmatrix}_{MS} 
    =
    \begin{bmatrix}
        T_{(k_A,\eta_A)}(\rho^-)
    \end{bmatrix}^{-1}
    \begin{bmatrix}
        \bar{\bar{I}} & \bar{\bar{O}} \\ \bar{\bar{C}}_{MS} & \bar{\bar{I}}
    \end{bmatrix}
    \begin{bmatrix}
        T_{(k_B,\eta_B)}(\rho^+)
    \end{bmatrix}
.\label{Wave_Metasurface}\end{equation}

\noindent{Note that if the metasurface is absent, $p_0$, $q_k$ and $r_k$ are all zero, so $\bar{\bar{C}}_{MS}=\bar{\bar{O}}$ and (\ref{Wave_Metasurface}) reduces to the wave matrix of a simple dielectric interface (\ref{Wave_Interface}).}

The discussion here has focused on metasurfaces with electric responses due to their ease of realization. Nevertheless, the procedure for deriving the wave matrix of metasurfaces is general. It can also be applied to metasurfaces with magnetic responses, or more complicated bianisotropic responses. This can be done by applying the relevant tangential boundary conditions.


\section{Synthesis of Field Transforming Devices}
\label{sec:synthesis}
In this section, the design procedure for cascaded metasurfaces that realize a specified field transformation is outlined. Strictly speaking, when a general field distribution is expressed as a summation of its azimuthal modes, an infinite number of orders should be taken into account. In reality, computational resources are limited so the summation must be truncated. Hence, (\ref{Ez_def}) is an approximation, and synthesizing the field transformation analytically is arduous.

Instead, an optimization process is adopted in this paper. As shown in Fig. \ref{fig2}, it is assumed that the sources are located in the central region of the metasurfaces. The first step of the design is to stipulate the field transformation that transforms the excitation field to some desired field in the outer region (outside the metasurfaces). 

\begin{figure}[!t]
\centerline{\includegraphics[width=7.5cm]{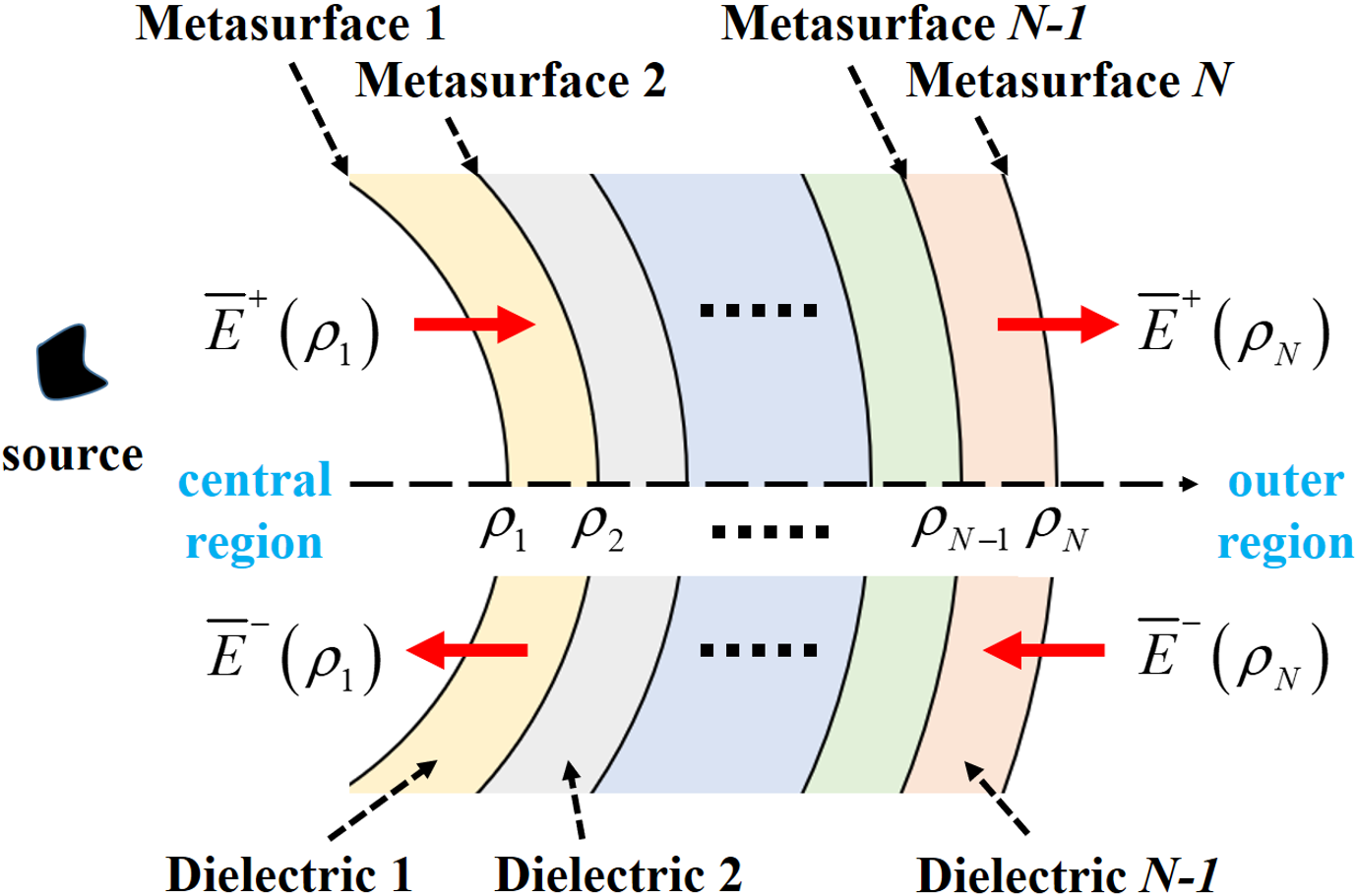}}
\caption{The cascaded structure, composed of $N$ layers of azimuthally-varying metasurfaces, is considered in the synthesis problem.}
\label{fig6}
\end{figure}

Next, we determine the numerical parameters used in the design, including the highest azimuthal order considered in the optimization process $M$, the number of cylindrical metasurface layers (denoted by $N$), as well as the highest order of variation $K$ for each admittance profile. The manner in which these parameters are chosen is discussed in detail in the Appendix. Furthermore, we also need to assign the radius of each cylindrical metasurface $\rho_1, \rho_2, ..., \rho_N$ and select the dielectric substrates between the layers. Note that in a practical realization, it is undesirable to have extremely small separations between metasurface layers. Otherwise, the metallic patterns on different metasurface layers may couple through evanescent waves or unwanted azimuthal modes. All these parameters are regarded as fixed throughout the synthesis process. As a result, the targeted structure is illustrated in Fig. \ref{fig6}.


In the synthesis process, the Fourier cosine and sine coefficients $p_0$, $q_k$ and $r_k$ of the admittance profile for every metasurface layer are optimized in order to realize the desired field outside the metasurfaces. The coefficients of each layer do not have to be the same, so there are $(2K+1)\times N$ variables in total. To initiate the first iteration of the optimization, $p_0$, $q_k$ and $r_k$ for each layer are selected randomly. The wave matrix of the entire cascaded structure is obtained from matrix multiplication as
\begin{equation}
\begin{split}
    \begin{bmatrix}
        W
    \end{bmatrix}_\text{all}
    = &
    \begin{bmatrix}
        W
    \end{bmatrix}_{MS_1}
    \begin{bmatrix}
        W
    \end{bmatrix}_{D_1}
    \begin{bmatrix}
        W
    \end{bmatrix}_{MS_2}
    \begin{bmatrix}
        W
    \end{bmatrix}_{D_2}
    \hdots 
    \\ &
    \hdots
    \begin{bmatrix}
        W
    \end{bmatrix}_{MS_{N-1}}
    \begin{bmatrix}
        W
    \end{bmatrix}_{D_{N-1}}
    \begin{bmatrix}
        W
    \end{bmatrix}_{MS_N}
.\end{split}
\label{eq_synthesis_1}\end{equation}
The wave matrices of the building blocks are found by substituting the physical parameters and Fourier coefficients into (\ref{Wave_Dielectric}), (\ref{Wave_Interface}), and (\ref{Wave_Metasurface}). We then convert this wave matrix $[W]_\text{all}$ to its corresponding $S$ matrix $[S]_\text{all}$ by employing (\ref{Wave2S}).

Now that the scattering matrix $[S]_\text{all}$ of the entire cascaded structure is known, the field distribution in both the central and outer regions can be calculated explicitly. In Fig. \ref{fig6}, nothing exists at infinity, so the electric field at $\rho_N$ simply propagates outward without any reflection. This indicates that $\bar{E}^-(\rho_N)$ is a zero vector. Multiplying out the block matrix equation of $S$ matrix definition (\ref{eq_S_Matrix_def}) yields
\begin{equation}
    \bar{\bar{c}}_1^- \bar{E}^-(\rho_1) = 
    \bar{\bar{S}}_{11\text{, all}} \cdot \bar{\bar{c}}_1^+ \bar{E}^+(\rho_1) 
\label{eq_synthesis_2}\end{equation}
\begin{equation}
    \bar{\bar{c}}_2^+ \bar{E}^+(\rho_N) = 
    \bar{\bar{S}}_{21\text{, all}} \cdot \bar{\bar{c}}_1^+ \bar{E}^+(\rho_1) 
.\label{eq_synthesis_3}\end{equation}

\noindent{The first equation (\ref{eq_synthesis_2}) can be solved by separating the incident field generated by the excitation from the field scattered by metasurfaces. Since it is assumed that all the sources are in the central region of the cascaded metasurfaces, the incident field generated by the sources must be outward propagating at $\rho_1$. Therefore,}
\begin{equation}
\begin{split}
    \bar{E}^+(\rho_1) &= \bar{E}_{inc}^+(\rho_1) + \bar{E}_{sca}^+(\rho_1) \\
    \bar{E}^-(\rho_1) &= \bar{E}_{sca}^-(\rho_1)
.\end{split}
\label{eq_synthesis_4}\end{equation}

\noindent{In (\ref{eq_synthesis_4}), $\bar{E}_{inc}^+(\rho_1)$ is already known because it is simply the field generated by the excitation when the metasurfaces are absent. On the other hand, the fields $\bar{E}_{sca}^+(\rho_1)$ and $\bar{E}_{sca}^-(\rho_1)$ together form standing waves since they are caused by scattering, and not by real sources. We can write}
\begin{equation}
\begin{split}
    \bar{E}_{sca}^+(\rho_1) &= 
    \big[ \beta_M H_M^{(2)}(k\rho_1), ..., \beta_{-M} H_{-M}^{(2)}(k\rho_1) \big]^T
    \\
    \bar{E}_{sca}^-(\rho_1) &= 
    \big[ \beta_M H_M^{(1)}(k\rho_1), ..., \beta_{-M} H_{-M}^{(1)}(k\rho_1) \big]^T
,\end{split}
\end{equation}
\noindent{where the $2M+1$ coefficients $\beta_M$, ..., $\beta_{-M}$ are the only unknowns in the $2M+1$ equations (\ref{eq_synthesis_2}). Accordingly, $\bar{E}^+(\rho_1)$ and $\bar{E}^-(\rho_1)$ can be solved explicitly from (\ref{eq_synthesis_2}). Substituting the calculated $\bar{E}^+(\rho_1)$ into (\ref{eq_synthesis_3}), the vector $\bar{E}^+(\rho_N)$ and thus the field in the outer region can be determined as well.}

The calculated $\bar{E}^+(\rho_N)$ for this current optimization iteration is then compared with the desired field outside the cascaded metasurfaces. A cost function $\mathcal{C}$ is defined to evaluate the error between the calculated field and the targeted field. It is minimized through the built-in $fmincon$ optimization algorithm in MATLAB. Useful example forms of the cost function will be shown in the next section. At the end of every iteration, the Fourier coefficients $p_0$, $q_k$ and $r_k$ of each metasurface layer are updated, brought into the wave matrix (\ref{eq_synthesis_1}), and a new $\bar{E}^+(\rho_N)$ is calculated until the optimization goal is achieved.



\section{Design Examples}
\label{sec:examples}
In order to verify the proposed wave matrix theory and demonstrate the field transforming ability of cylindrical metasurfaces, three metasurface designs are synthesized, simulated and discussed in this section. The full-wave simulations of the targeted structure illustrated in Fig. \ref{fig6} are conducted using the COMSOL Multiphysics 2D finite element electromagnetic solver. The metasurface admittance (\ref{eq_Y_profile}) is modeled with surface current densities that are dependent on electric fields. Without loss of generality, lossless devices are considered in this paper; therefore, the metasurface admittances are all purely imaginary. The central region (where $\rho<\rho_1$) and the outer region ($\rho>\rho_N$) are taken to be free-space. The outermost edge of the computational domain (at some $\rho \gg \rho_N$) is terminated by a Perfect Matched Layer (PML).

\subsection{Case 1: Azimuthal Mode Converter}
The first device we are interested in is an azimuthal mode converter that transform a $m=0$ ($0^{th}$ azimuthal mode) excitation field to a $m=1$ ($1^{st}$ azimuthal mode) field outside the cascaded metasurfaces. Three metasurfaces ($N=3$) with radii $\rho_1 = 1.6\lambda$, $\rho_2 = 2.2\lambda$, and $\rho_3 = 2.8\lambda$ are assumed in the azimuthal mode converter design ($\lambda$ stands for the operating wavelength). The dielectric spacers between the metasurfaces are chosen to be air. In addition, the order of spatial variation on each metasurface is chosen to be $K=5$.

The source is assumed to be an electric line current with unit amplitude located at the origin, which produces an $m=0$ excitation. In this case, the incident field at $\rho_1$ can be found from applying Ampere's law around the line source,
\begin{equation}
    E_{z,inc}(\rho_1, \phi) = -\frac{k_0^2}{4\omega\varepsilon_0}H_0^{(2)}(k_0\rho_1)
.\label{eq_caseA_1}\end{equation}

\noindent{The incident field vector $\bar{E}_{inc}^+(\rho_1)$ in (\ref{eq_synthesis_4}) can be derived by decomposing (\ref{eq_caseA_1}) into azimuthal modes. Specifically, only the $m=0$ component $\mathcal{E}^+_{0,inc}(\rho_1)$ is nonzero in $\bar{E}_{inc}^+(\rho_1)$:}
\begin{equation}
    \mathcal{E}^+_{0,inc}(\rho_1) = -\frac{k_0^2}{4\omega\varepsilon_0}H_0^{(2)}(k_0\rho_1)
.\end{equation}
\noindent{With the expression of the vector $\bar{E}_{inc}^+(\rho_1)$, the optimization process proposed in section IV can be performed. Outside the device an $m=1$ field is desired, so $E_z(\rho_3,\phi)$ should be of the form $H_1^{(2)}(k_0\rho_3) e^{-j\phi}$. Consequently, the following cost function can be defined [\ref{my_APS}],}
\begin{equation}
    \mathcal{C} = \Big{[} 
    \frac{\text{Power of } H_1^{(2)} \text{ mode in the outer region}}{\text{Total power in the outer region}} - 1
    \Big{]}^2
.\label{eq_caseA_2}\end{equation}

\begin{figure}[!t]
    \centering
    \subfloat[]{\includegraphics[width=8.8cm]{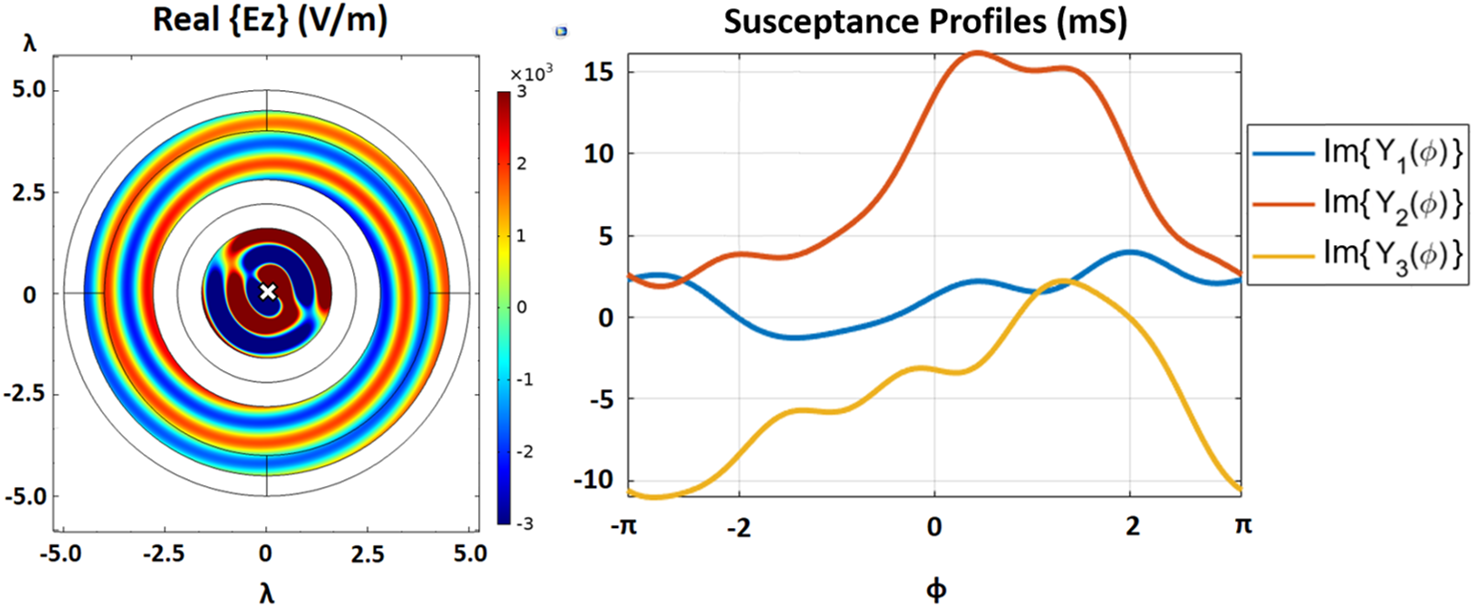}}
    \\
    \subfloat[]{\includegraphics[width=8.8cm]{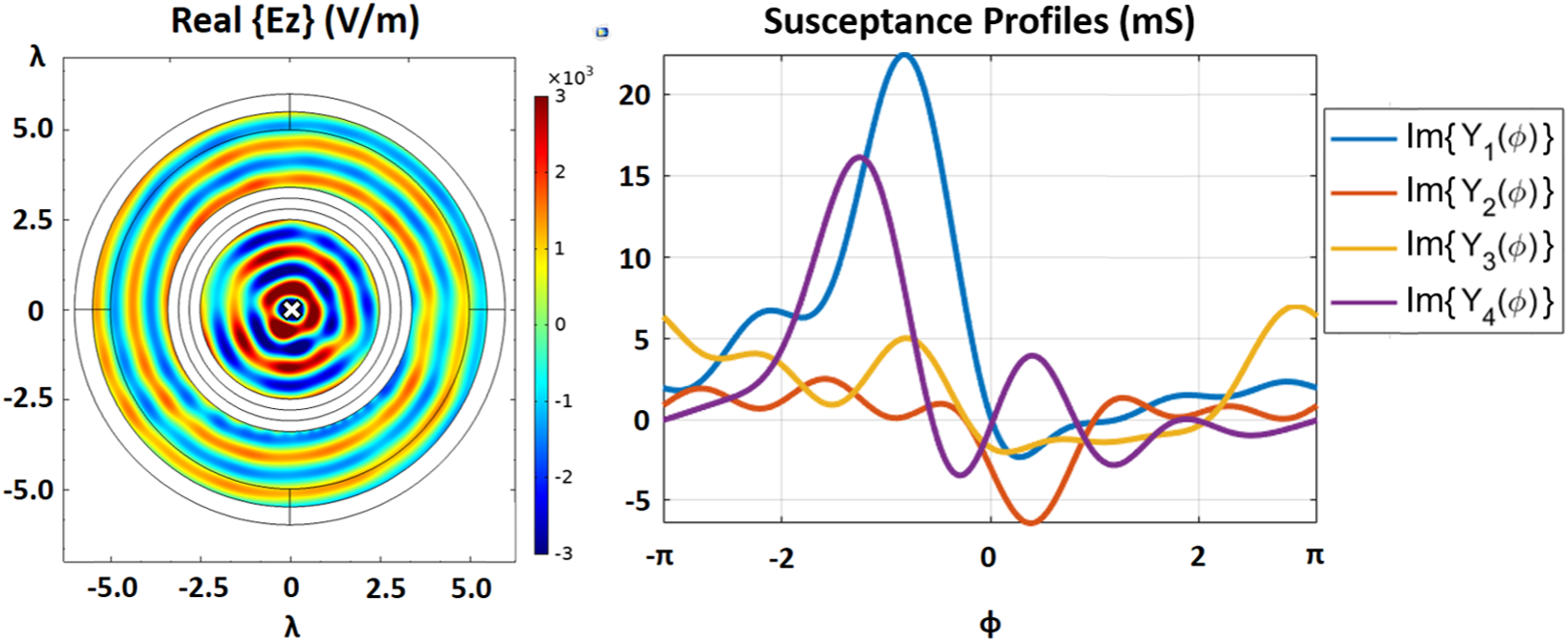}}
    \caption{The designed azimuthal mode converter which converts the $m=0$ line current excitation to $m=1$ output field. The cross symbols in the field plot (from COMSOL simulation) indicates the location of the excitation line current. Susceptance (imaginary part of admittance) profiles of the metasurfaces are also shown. (a) The case utilizing reflection waves within the central region. (b) The case with reduced reflection.}
\label{fig_caseA}\end{figure}

\noindent{In (\ref{eq_caseA_2}), a restriction is placed only on the field outside the cascaded metasurfaces, which means that reflections within the central region are allowed. A total of $2M+1=31$ azimuthal orders are considered in all regions during the optimization process. The admittance profiles (purely imaginary) obtained through optimization are simulated in COMSOL, resulting in the field plot shown in Fig. \ref{fig_caseA} (a). The power in the $m = 1$ mode constitutes 99\% of the total power in the outer region, and the specified field transformation is clearly accomplished.}

The azimuthal mode converter can also be designed to minimize reflections in the central region. Since the field transformation has the additional constraint of being reflectionless, we choose $N=4$ (four metasurfaces) and $K=5$. The radii of the metasurfaces are $\rho_1=2.5\lambda$, $\rho_2=2.8\lambda$, $\rho_3=3.1\lambda$, $\rho_4=3.4\lambda$, and air is used for the dielectric spacers in between. The cost function is now defined as [\ref{my_APS}],
\begin{equation}
\begin{split}
    \mathcal{C} &=
    a\Big{[} 
    \frac{\text{Power of } H_1^{(2)} \text{ mode in the outer region}}{\text{Total power in the outer region}} - 1
    \Big{]}^2\\
    &+
    b\Big{[} 
    \frac{\text{Power of } H_0^{(2)} \text{ mode in the central region}}{\text{Total power in the central region}} - 1
    \Big{]}^2
.\end{split}
\label{eq_caseA_3}\end{equation}

\noindent{In this example $2M+1=41$, and the ratio of $a/b$ is taken to be 10. The optimized admittance profiles (purely imaginary), together with the full-wave simulation result, are displayed in Fig. \ref{fig_caseA} (b). The power in the $m = 1$ mode constitutes 97\% of the total power in the outer region, while the power in the $m = 0$ mode constitutes 87\% of the total power in the central region. Although the field outside the device is not as perfect as in the previous case, reflections are significantly reduced in the central region. This example helps demonstrate the versatility of the proposed wave matrix framework.}

\begin{figure}[!t]
\centering
\subfloat[\footnotesize\label{fig_caseAa}]{%
    \includegraphics[width=0.5\linewidth]{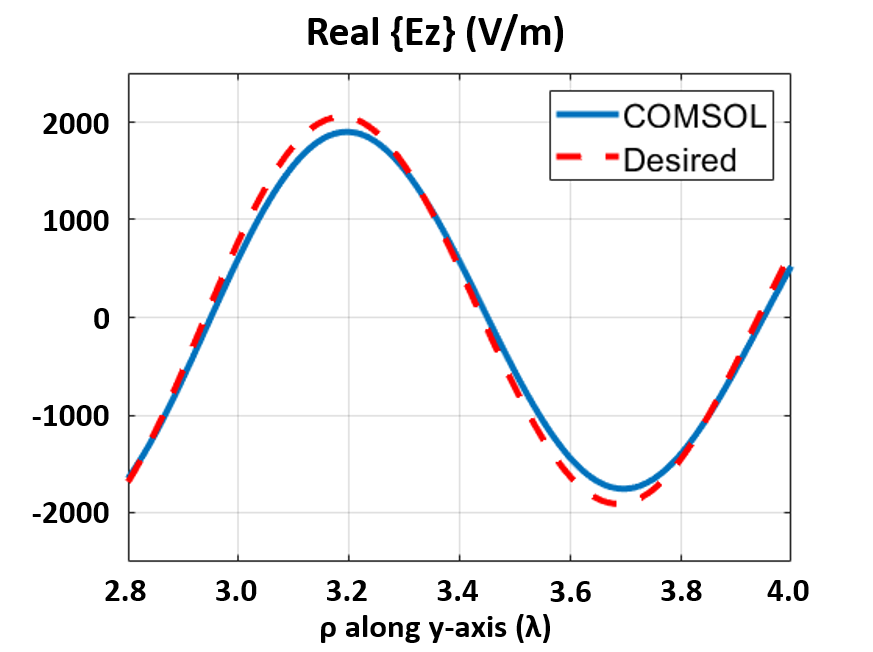}}
\subfloat[\footnotesize\label{fig_caseAb}]{%
    \includegraphics[width=0.5\linewidth]{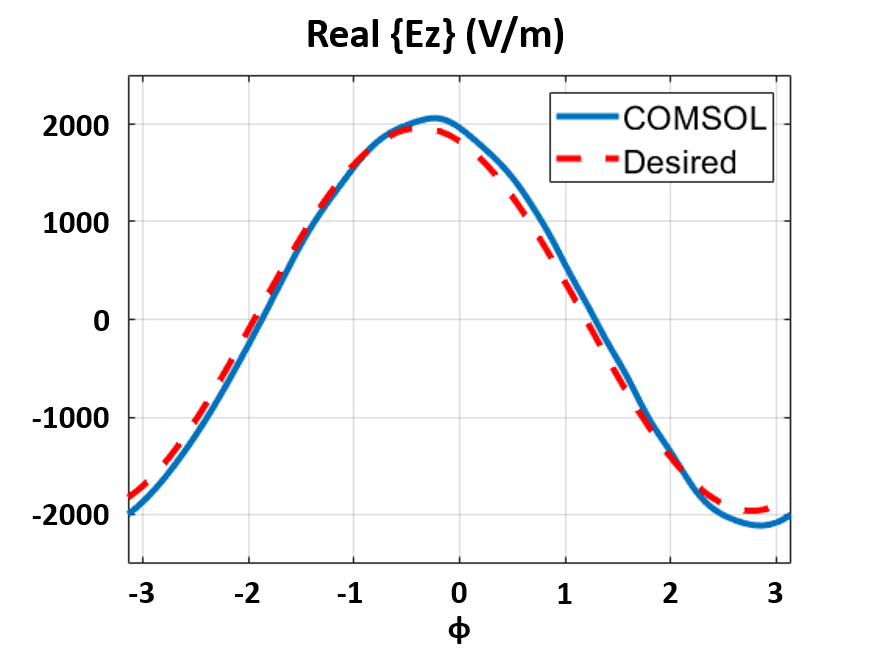}}
\caption{Real part of the electric field (time snapshot) for the azimuthal mode converter discussed in Fig. \ref{fig_caseA} (a). (a) Along the $y$-axis. (b) Along a cylindrical surface at $\rho=3.5\lambda$. Both simulated (COMSOL) and desired fields are shown.}
\label{fig_caseA_quant1} 
\end{figure}

\begin{figure}[!t]
\centering
\subfloat[\footnotesize\label{fig_caseAc}]{%
    \includegraphics[width=0.5\linewidth]{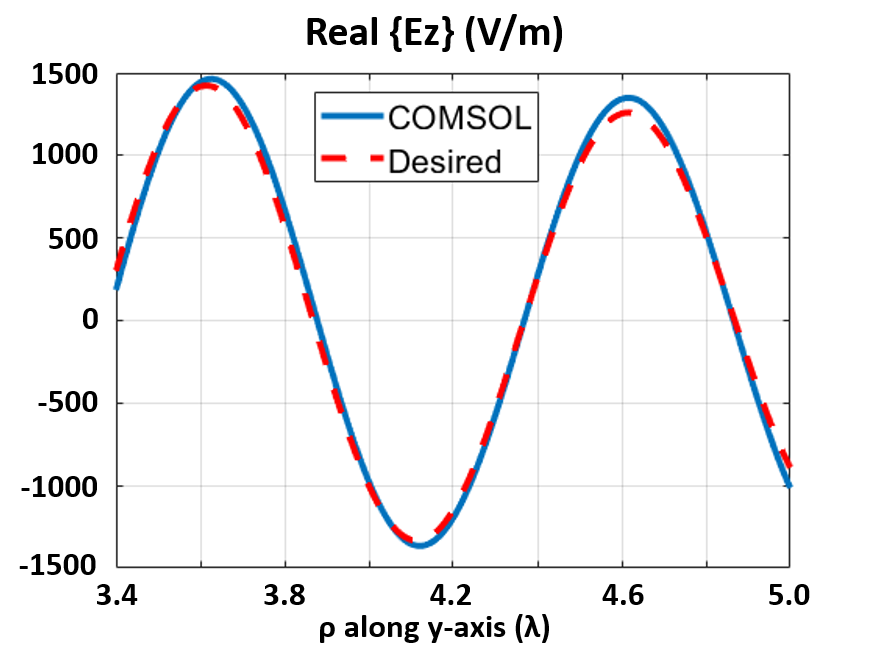}}
\subfloat[\footnotesize\label{fig_caseAad}]{%
    \includegraphics[width=0.5\linewidth]{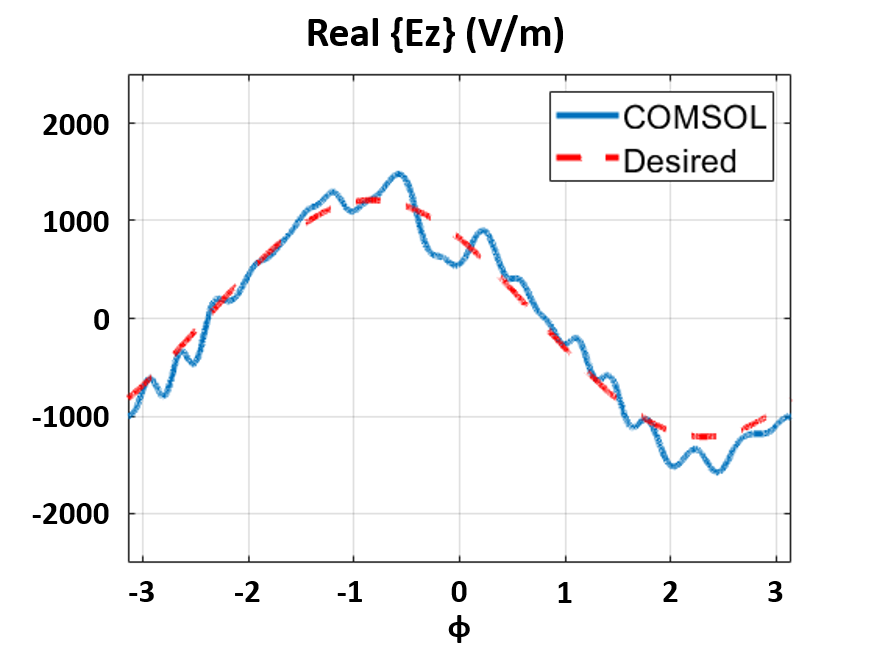}}
\caption{Real part of the electric field (time snapshot) for the azimuthal mode converter discussed in Fig. \ref{fig_caseA} (b). (a) Along the $y$-axis. (b) Along a cylindrical surface at $\rho=5.0\lambda$. Both simulated (COMSOL) and desired fields are shown.}
\label{fig_caseA_quant2} 
\end{figure}

To better understand the performances of the azimuthal mode converters, cross-sectional views of the simulated and desired electric fields (real parts) in the outer region are plotted in Fig. \ref{fig_caseA_quant1} and \ref{fig_caseA_quant2}. In Fig. \ref{fig_caseA_quant1} (a) and \ref{fig_caseA_quant2} (a), the fields are plotted along the $y$-axis. Close agreement is observed between the COMSOL simulation results and the desired fields. In Fig. \ref{fig_caseA_quant1} (b) and \ref{fig_caseA_quant2} (b), the fields are plotted along a cylindrical surface. As can be seen, the real part of the electric field undergoes one complete sinusoidal variation from $\phi=-\pi$ to $+\pi$, as expected for an $m=1$ cylindrical mode. Note that there are some discrepancies between the simulated and desired fields in Fig. \ref{fig_caseA_quant2} (b). In this design, the performance in the outer region is slightly sacrificed to reduce reflections within the central region.

Recently, light beams with orbital angular momenta (OAM) have been used in particle trapping and manipulation, as well as high-speed communication [\ref{Li(2)_2019}, \ref{He_1995}-\ref{Wang_2012}]. The azimuthal mode converters reported here demonstrate the possibility of generating OAM beams using relatively simple structures instead of bulky antenna arrays [\ref{Song_2021}].

\subsection{Case 2: Illusion Device}
The next example proposed in this paper is an illusion device. We again excite the device using a line current source, but design the metasurfaces so that the source appears as if it was displaced in space. Four metasurfaces ($N=4$) with radii $\rho_1=2.0\lambda$, $\rho_2 = 2.3\lambda$, $\rho_3 = 2.6\lambda$, and $\rho_4 = 2.9\lambda$ are used to synthesize the device. The order of spatial variation $K$ on each metasurface is set to be 4. Again, the dielectric spacers are all chosen to be air.  

The electric field at ($\rho$,$\phi$) generated by an off-centered line current located at ($\rho'$, $\phi'$) is given by [\ref{Harrington}]
\begin{equation}
    E_{z,inc}(\rho,\phi) = 
    -\frac{k_0^2}{4\omega\varepsilon} \sum_m H_m^{(2)}(k\rho') J_m(k\rho) e^{-jm(\phi-\phi')} 
\label{eq_caseB_1}\end{equation}
\noindent{when $\rho \leq \rho'$, and}

\begin{equation}
    E_{z,inc}(\rho,\phi) = 
    -\frac{k_0^2}{4\omega\varepsilon} \sum_m J_m(k\rho') H_m^{(2)}(k\rho) e^{-jm(\phi-\phi')} \\
\label{eq_caseB_2}\end{equation}
\noindent{when $\rho \geq \rho'$. Here the device is excited by a line source at $0.5\lambda$ to the right of the origin ($\rho'=0.5\lambda$, $\phi'=0$). Therefore, equation (\ref{eq_caseB_2}) is used to determine the incident field at $\rho_1$. The azimuthal components of the incident field at $\rho_1$ are}
\begin{equation}
\begin{split}
    \mathcal{E}_{m,inc}^+(\rho_1)
    & = -\frac{k_0^2}{4\omega\varepsilon_0} J_m(\pi) H_m^{(2)}(k_0\rho_1) \\
    & = \alpha_{m,inc}^+ H_m^{(2)}(k_0\rho_1)
.\end{split}
\label{eq_caseB_3}\end{equation}

\noindent{On the other hand, the illusion is set to be at $0.5\lambda$ to the left of the origin ($\rho'=0.5\lambda$, $\phi'=\pi$). The targeted output field at $\rho_4$ can also be expressed using (\ref{eq_caseB_2}). Ideally, the azimuthal components are}
\begin{equation}
\begin{split}
    \mathcal{E}_{m,ideal}^+(\rho_4)
    & = -\frac{k_0^2}{4\omega\varepsilon_0} (-1)^m J_m(\pi) H_m^{(2)}(k_0\rho_4) \\
    & = \alpha_{m,ideal}^+ H_m^{(2)}(k_0\rho_4)
.\end{split}
\label{eq_caseB_4}\end{equation}

\noindent{In order to define the cost function, we denote the calculated azimuthal components at each iteration as}
\begin{equation}
    \mathcal{E}_{m,cal}^+(\rho_4)
     = \alpha_{m,cal}^+ H_m^{(2)}(k_0\rho_4)
.\label{eq_caseB_5}\end{equation}

\noindent{In this case, a possible cost function to be minimized can be}
\begin{equation}
    \mathcal{C} = \sum_m \big| \hat{\alpha}_{m,cal}^+ - \hat{\alpha}_{m,ideal}^+ \big|^2
,\label{eq_caseB_6}\end{equation}

\noindent{where the quantities $\hat{\alpha}_{m,cal}^+$ and $\hat{\alpha}_{m,ideal}^+$ are normalized wave amplitudes:}
\begin{equation}
    \hat{\alpha}_{m,cal}^+ = \frac{\alpha_{m,cal}^+}{ \sqrt{ \sum_m |\alpha_{m,cal}|^2 } }
    \text{, } 
    \hat{\alpha}_{m,ideal}^+ = \frac{\alpha_{m,ideal}^+}{ \sqrt{ \sum_m |\alpha_{m,ideal}|^2 } }
.\label{eq_caseB_7}\end{equation}

\noindent{The line current is an impressed source, and the cascaded metasurface structure appears as a load to the source. For each optimization iteration, the load is changed, so the power transmitted to the outer region also changes. Therefore, there could be an overall magnitude difference between the calculated wave amplitudes $\alpha^+_{m,cal}$ and the ideal ones $\alpha^+_{m,ideal}$. Hence, the wave amplitudes are normalized to remove this magnitude difference. In essence, this cost function demands that the optimized azimuthal components be as close to the ideal ones as possible.}

\begin{figure}[!t]
\centerline{\includegraphics[width=8.5cm]{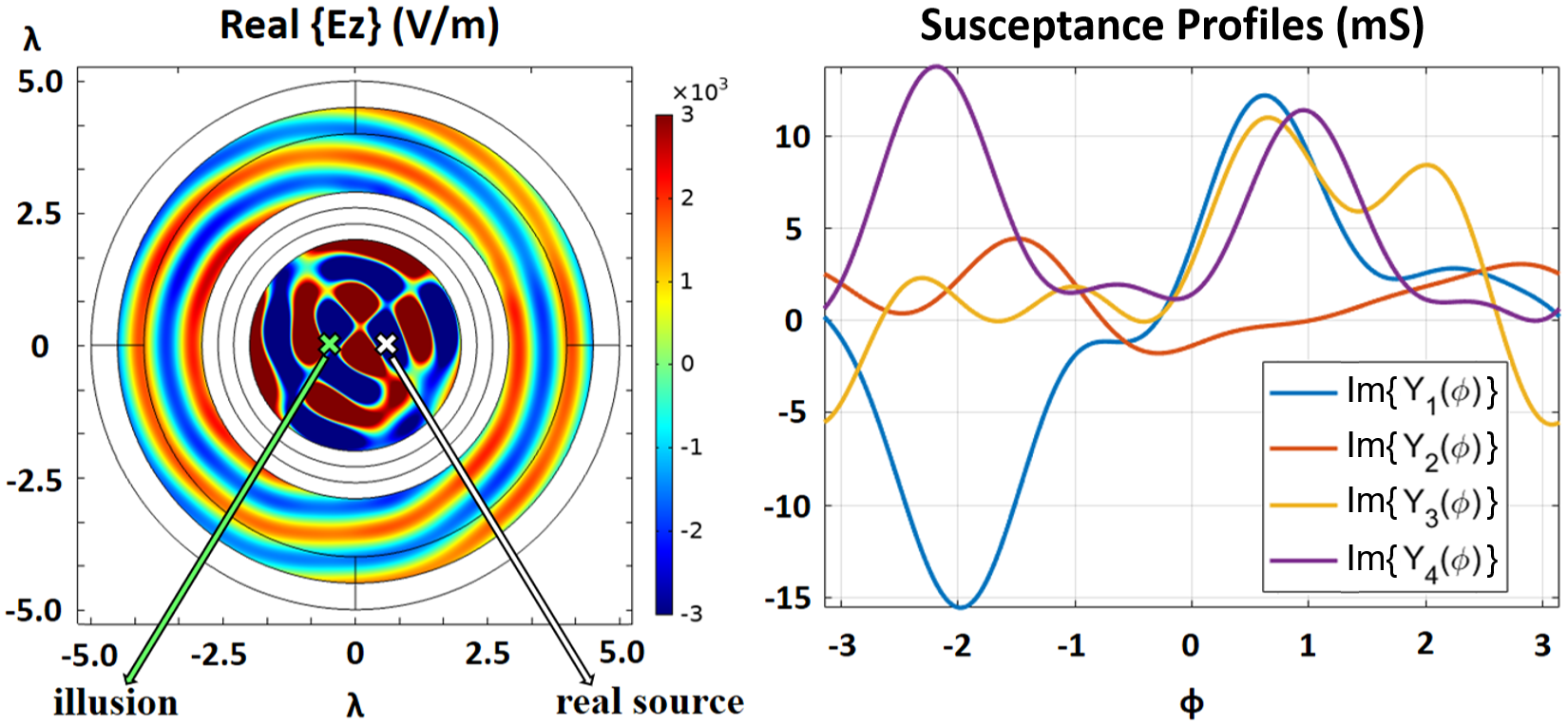}}
\caption{The designed illusion device tailors electromagnetic fields so that the source appears displaced in space. The real source is marked as the white cross symbol, and the illusion of the source is represented by the green one.}
\label{fig_caseB}
\end{figure}

\begin{figure}[!t]
\centering
\subfloat[\footnotesize\label{fig_caseBa}]{%
    \includegraphics[width=0.5\linewidth]{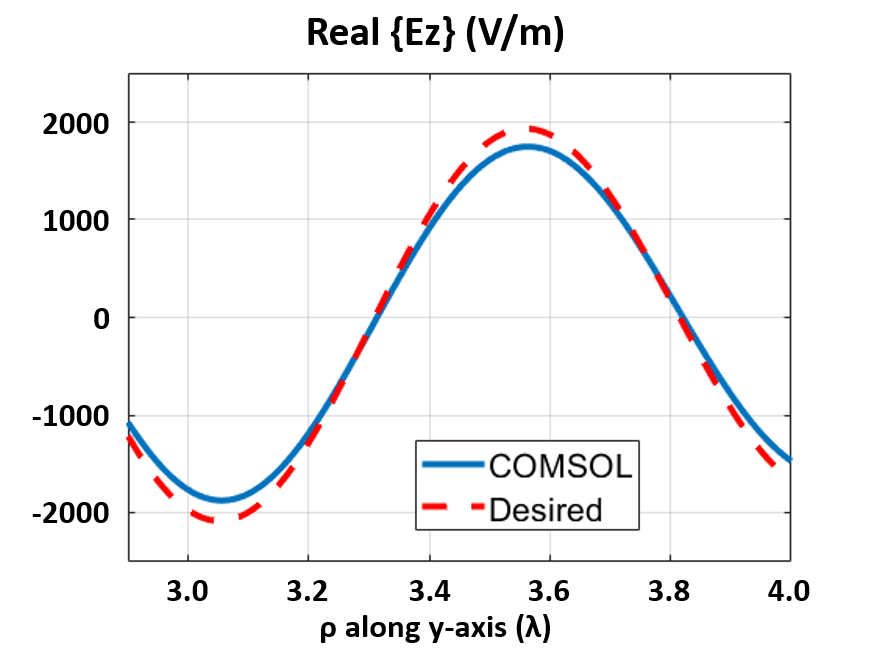}}
\subfloat[\footnotesize\label{fig_caseBb}]{%
    \includegraphics[width=0.5\linewidth]{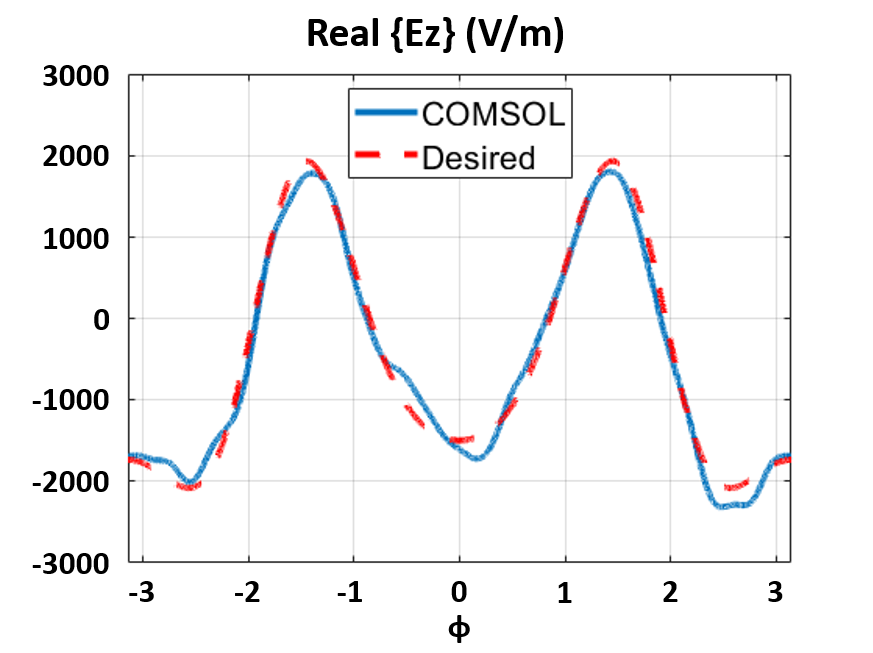}}
\caption{Real part of the electric field (time snapshot) for the illusion device. (a) Along the $y$-axis. (b) Along a cylindrical surface at $\rho=3.5\lambda$. Both simulated (COMSOL) and desired fields are shown.}
\label{fig_caseB_quant} 
\end{figure}

Employing the cost function (\ref{eq_caseB_6}), the optimized admittance profiles (purely imaginary) of the four metasurfaces are plotted in Fig. \ref{fig_caseB}, resulting in a value of $\mathcal{C} = 0.0105$. The total number of azimuthal modes considered in the optimization is $2M+1=49$. These admittance profiles are again simulated in COMSOL, yielding the field plot in Fig. \ref{fig_caseB}. From outside the cascaded metasurfaces it seems that the field is originating from ($\rho'=0.5\lambda$, $\phi'=\pi$), while the actual source is located at ($\rho'=0.5\lambda$, $\phi'=0$). As in the previous examples, cross-sectional views of the simulated and desired electric fields in the outer region are plotted in Fig. \ref{fig_caseB_quant}. Close agreement is again observed, confirming the accuracy of the proposed design method. This example illustrates possible applications to stealth technologies, where illusions or radar cross-section reductions are of concern.

\subsection{Case 3: Multi-functional Metasurface}
Earlier devices constructed from passive cylindrical metasurfaces performed single-input single-output operations. That is, they could only transform a single, fixed input field to a corresponding output field. In this paper, the final device presented is a multi-functional metasurface. The multi-functional metasurface performs differently when it is excited by different sources. For instance, when the device is excited by a line current source located at ($\rho'=1.5\lambda$, $\phi'=0$), it generates an $m=0$ output field. However, when the same device is fed by a source at ($\rho'=1.5\lambda$, $\phi'=\pi$) , it produces an $m=2$ output field. For convenience, the line current sources at ($\rho'=1.5\lambda$, $\phi'=0$) and ($\rho'=1.5\lambda$, $\phi'=\pi$) are named source 1 and source 2, respectively.

\begin{figure}[!t]
    \centering
    \subfloat[]{\includegraphics[width=8.5cm]{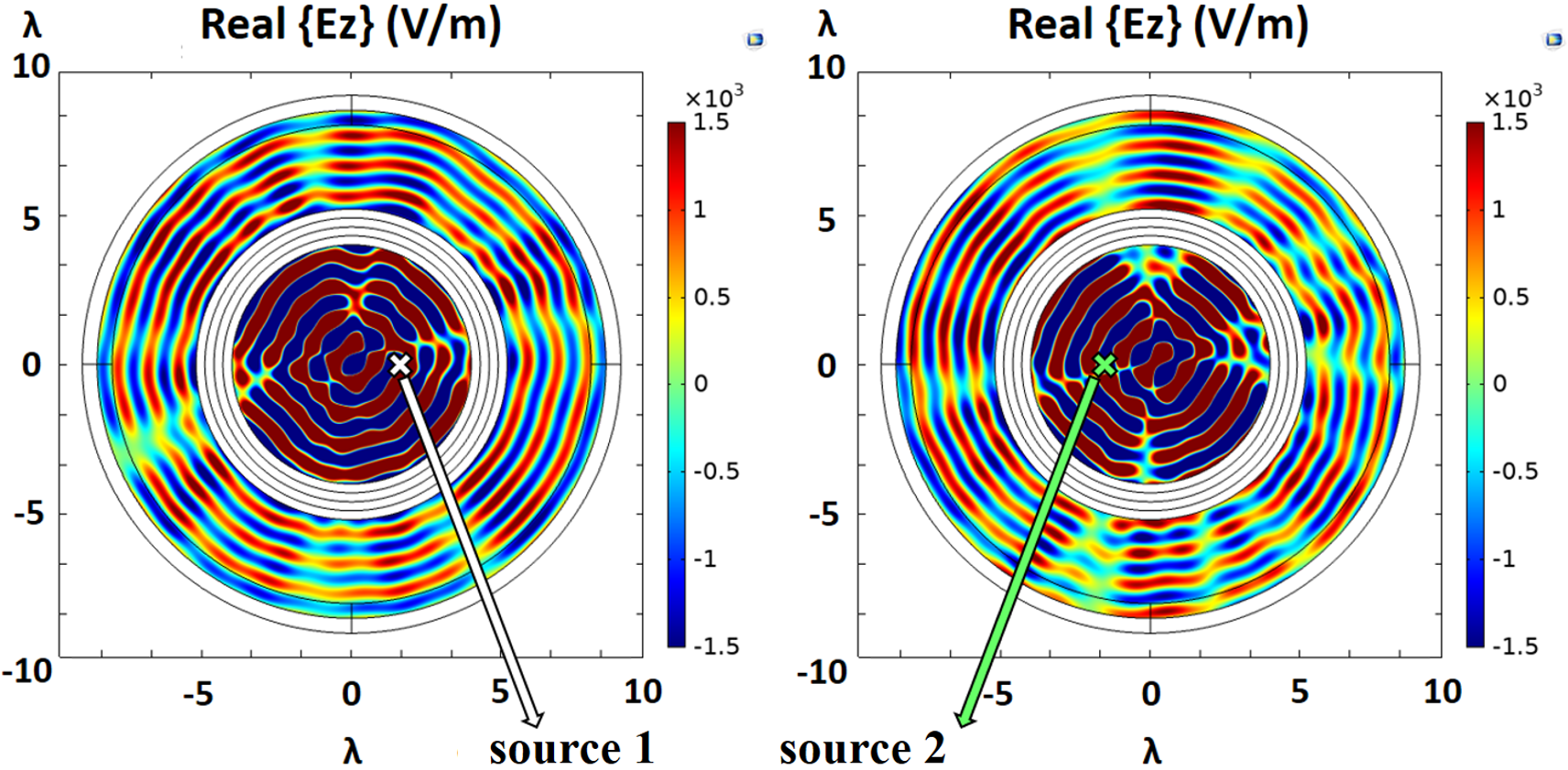}}
    \\
    \subfloat[]{\includegraphics[width=6.4cm]{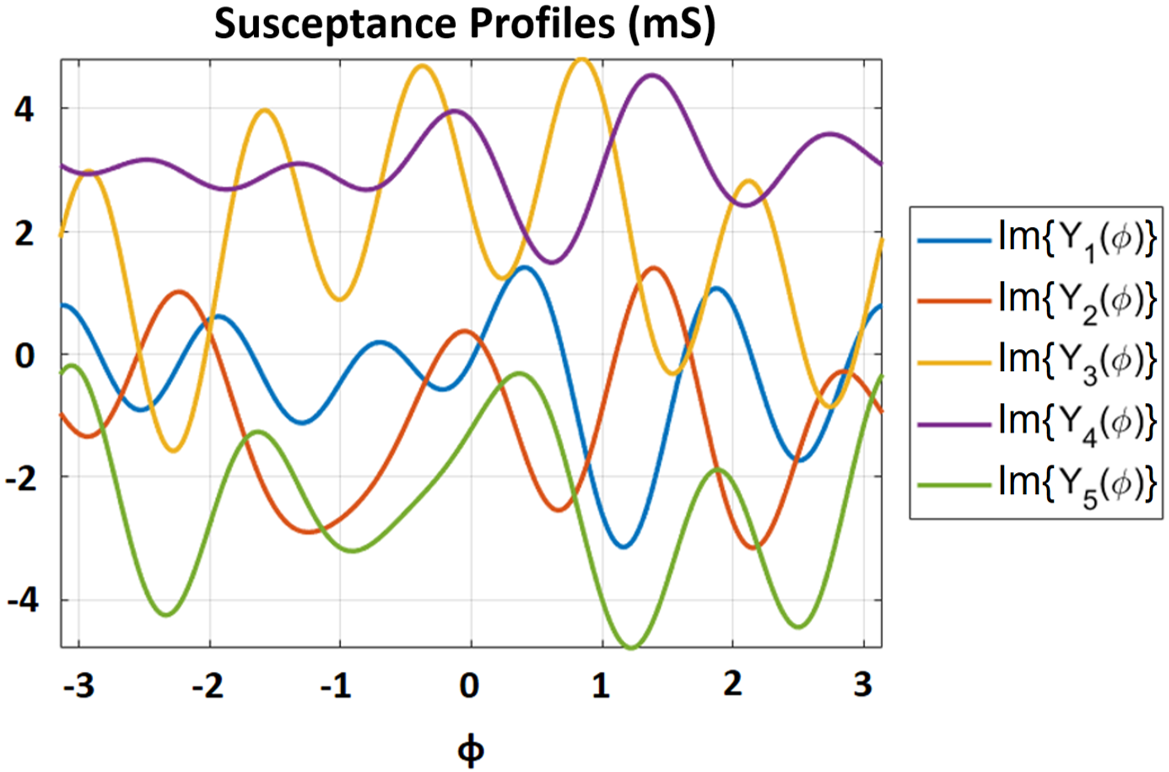}}
    \caption{An example of multi-functional metasurfaces. (a) The figure on the left displays the field plot when the device is fed by source 1. The figure on the right shows the field plot when source 2 is used as the excitation. An $m=2$ dependence in the output field is clear. (b) The optimized susceptance (imaginary part of admittance) profiles for all five metasurface layers.}
\label{fig_caseC}\end{figure}

To synthesize this device, five metasurfaces are used ($N=5$), with order of variation $K=5$. The radii are chosen to be $\rho_1 = 4.0\lambda$, $\rho_2 = 4.3\lambda$, $\rho_3 = 4.6\lambda$, $\rho_4 = 4.9\lambda$, and $\rho_5 = 5.2\lambda$. Finally, all the dielectric spacers are set to be air. Before introducing the cost function, we define two power ratios:
\begin{equation}
    R_1 =
    \frac{\text{Power of } H_0^{(2)} \text{ mode in the outer region}}{\text{Total power in the outer region}}
\label{eq_caseC_1}\end{equation}
\noindent{in which both the numerator and denominator are evaluated when only source 1 is excited. Similarly,}
\begin{equation}
    R_2 =
    \frac{\text{Power of } H_2^{(2)} \text{ mode in the outer region}}{\text{Total power in the outer region}}
\label{eq_caseC_2}\end{equation}
\noindent{in which both the numerator and denominator are evaluated when only source 2 is excited. The cost function takes the form:}
\begin{equation}
    \mathcal{C} = a(R_1 - 1)^2 + b(R_2 - 1)^2
.\label{eq_caseC_3}\end{equation}

\noindent{In this example, both $a$ and $b$ are set to be 1 so that the two cases have equal weightings.}

Based on the proposed cost function (\ref{eq_caseC_3}), optimization in MATLAB and simulation in COMSOL are carried out. A total of $2M+1=61$ azimuthal modes are included in the optimization process. The electric field plots, along with the required purely imaginary admittance profiles, are shown in Fig. \ref{fig_caseC} (a) and (b) respectively. Under the excitation of source 1, the power in the $m = 0$ mode constitutes 77\% of the total power in the outer region. Under the excitation of source 2, the power in the $m = 2$ mode constitutes 79\% of the total power in the outer region. The synthesized multi-functional device successfully transforms source 1 to an $m=0$ field. It also transforms source 2 to an $m=2$ field. 

\begin{figure}[!t]
\centering
\subfloat[\footnotesize\label{fig_caseCa}]{%
    \includegraphics[width=0.5\linewidth]{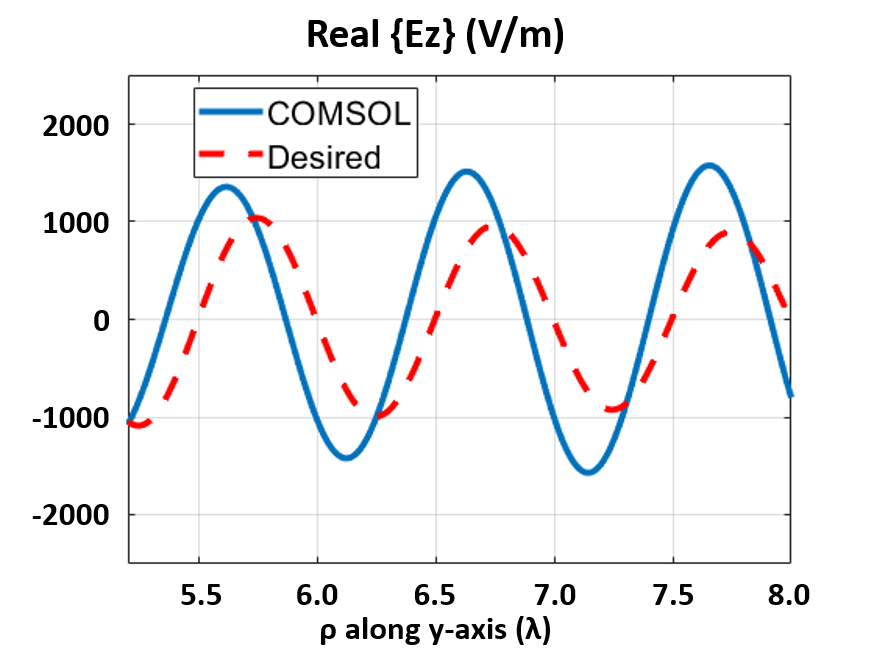}}
\subfloat[\footnotesize\label{fig_caseCb}]{%
    \includegraphics[width=0.5\linewidth]{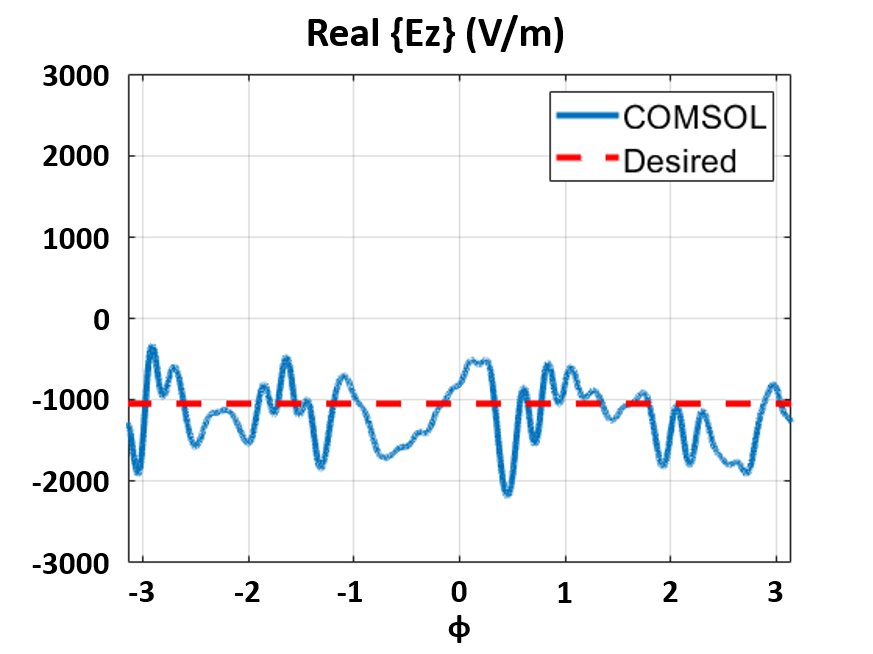}}
\vfill
\subfloat[\footnotesize\label{fig_caseCc}]{%
    \includegraphics[width=0.5\linewidth]{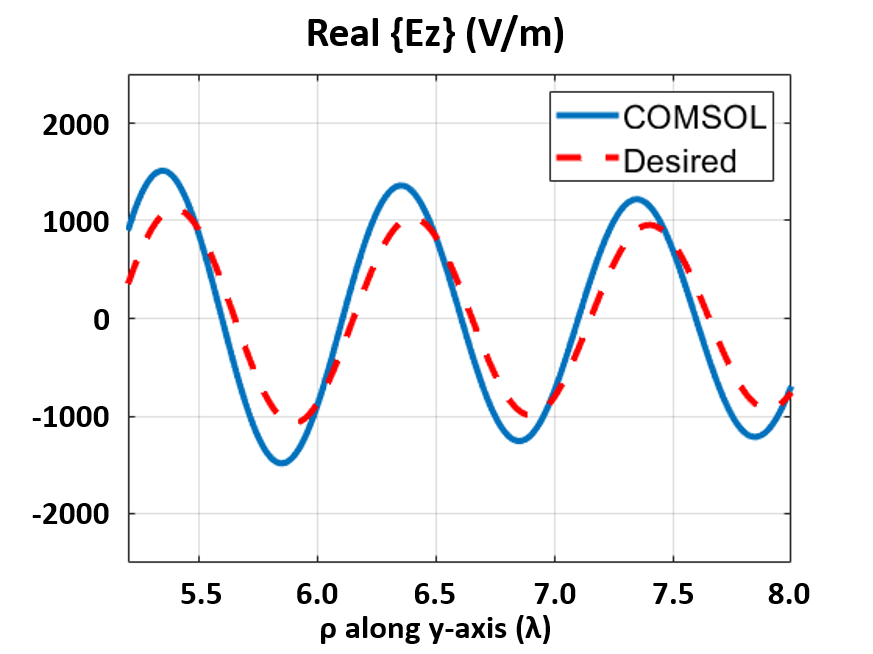}}
\subfloat[\footnotesize\label{fig_caseCd}]{%
    \includegraphics[width=0.5\linewidth]{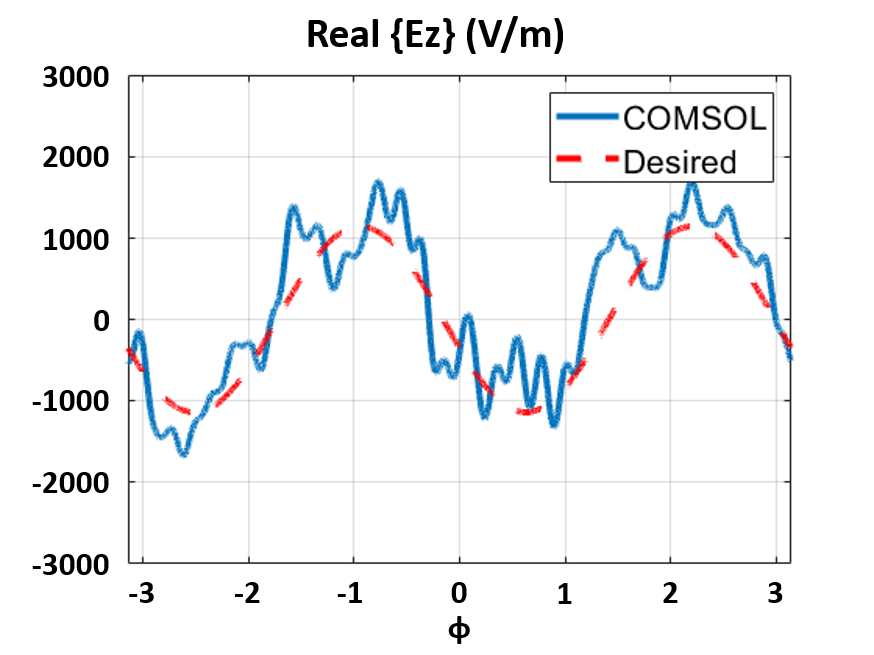}}
\caption{Real part of the electric field (time snapshot) for the multi-functional metasurface. When fed by source 1: (a) Along the $y$-axis. (b) Along a cylindrical surface at $\rho=5.2\lambda$. When fed by source 2: (c) Along the $y$-axis. (d) Along a cylindrical surface at $\rho=5.2\lambda$.}
\label{fig_caseC_quant} 
\end{figure}

Cross-sectional views of the electric field when the device is excited by source 1 are shown in Fig. \ref{fig_caseC_quant} (a) and (b). Due to the larger portion of undesired ($m \neq 0$) modes, the simulated field does not as closely match the desired field as in the previous examples. Although in Fig. \ref{fig_caseC_quant} (b) there are more oscillations in the simulated field, the averaged value still agrees with the desired field with an $m=0$ azimuthal dependence. Similarly, cross-sectional views of the electric field when the device is excited by source 2 are shown in Fig. \ref{fig_caseC_quant} (c) and (d). As can be seen in Fig.\ref{fig_caseC_quant} (d), the real part of the electric field undergoes two complete sinusoidal variation from $\phi = -\pi$ to $+\pi$, as expected for an $m=2$ cylindrical mode.

This device may be a promising antenna feed for multi-channel orbital angular momentum-based links, and Multi-Input Multi-Output (MIMO) antennas in general. More importantly, it demonstrates the power and the feasibility of the proposed framework to realize arbitrary field transformation.


\section{Guidelines on Practical Realization}
\label{sec:realization}
Although a two-dimensional scenario is considered in this paper, the concepts can easily be applied to realistic devices. To this end, several important suggestions on actual implementation are provided in this section.

\begin{figure}[!t]
\centerline{\includegraphics[width=8.5cm]{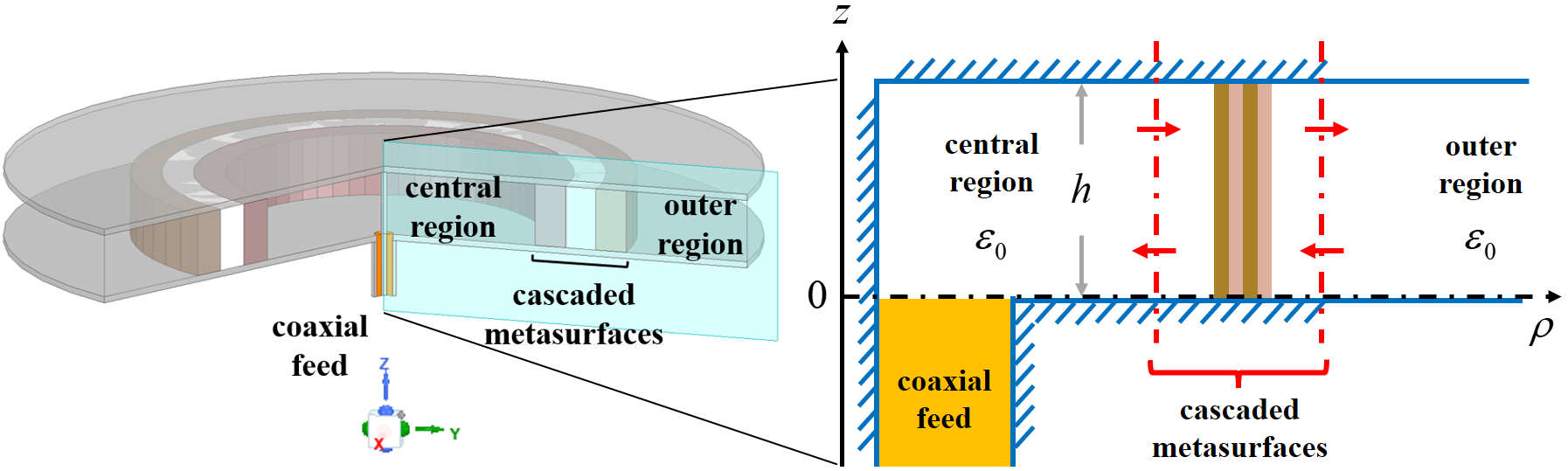}}
\caption{A possible structure that can realize the cylindrical metasurfaces considered in this paper. It is assumed in this diagram that the feed is located at the center. (a) The three-dimensional view. (b) The cross-sectional view on a vertical plane.}
\label{realization}
\end{figure}

A possible structure that can realize the azimuthally-varying, cascaded, cylindrical metasurfaces considered in this paper is proposed in Fig. \ref{realization}. The cascaded cylindrical metasurfaces are inserted within a radial or a parallel-plate waveguide with large conducting plates located at $z = 0$ and $z = h$. The height of the radial waveguide, $h$, should be less than half a wavelength $\lambda$, in order to ensure that all the propagating fields inside the waveguide are $z$-independent. Therefore, only waves with vanishing $k_z$ are allowed to propagate radially. Under this restriction, solving for the propagating fields within the waveguide reduces to a two-dimensional problem, as discussed in this paper.

The concentric cylindrical metasurfaces, as well as the radial waveguide, are excited by a coaxial cable feed oriented in the $z$-direction, which mimics a line current source. For simplicity, it is assumed that the center conductor of the coaxial feed touches the upper plate of the waveguide. In Fig. \ref{realization}, the coaxial cable feed is depicted at the center of the cylindrical metasurfaces. However, this is not necessary, as we have already seen in Parts B and C of the previous section. In contrast to the ideal current source case, the scattering properties of the coaxial feed need to be characterized and accounted for in the design procedure. This characterization may be performed by applying the mode-matching technique [\ref{Heebl_2016}, \ref{Shen_1999}].

To realize the concentric metasurfaces, the continuous admittance profile of each cylindrical metasurface (\ref{eq_Y_profile}) is discretized into unit cells. The capacitive admittance values of the unit cells can be implemented as metallic patches or interdigitated capacitors on a flexible substrate such as 0.203-mm Rogers 4003C [\ref{Raeker_2016}]. Inductive values can be realized using metallic strips or meander lines. However, excessively thin strips or long meander lines may introduce additional ohmic loss. Therefore, it is recommended to avoid highly inductive admittances in practical realizations.

To properly select the metallic patterning for each unit cell, the admittance values of metallic patterns needs to be extracted. In order to account for the curvature of cylindrical metasurfaces, it can be assumed that the target unit cell repeats itself in the $\phi$ direction, and forms a full cylindrical surface [\ref{Li(2)_2019}]. This local periodicity assumption, widely applied in the design of planar metasurfaces, is sufficient when the admittance profiles (\ref{eq_Y_profile}) to be realized are not highly oscillatory or do not contain abrupt changes. The admittance variation can be controlled by simply limiting the highest azimuthal order of admittance variation ($K$). The admittance of a patterned unit cell can be extracted by comparing its scattering parameters to those of an idealized admittance sheet [\ref{Li(2)_2019}].

With all the steps suggested in this section, the metallic patterning of the unit cells can be designed, printed on flexible substrates, and wrapped around dielectric spacers into the required cylindrical shape. In summary, although practical realization is not the focus of this paper, the proposed formulation has provided a straightforward path toward building realistic devices.


\section{Conclusion}
\label{sec:conclusion}

Field synthesis and transformation with cascaded cylindrical metasurfaces have been investigated in literature using idealized bianisotropic boundary conditions. However, physical realizations continue to be a challenge due to the requirements imposed by these idealized boundary conditions, such as extremely close metasurface separations as well as the need for perfect conducting baffles. These intricate and costly structures are needed to prevent higher order azimuthal mode coupling from propagating between metasurface layers. In this paper, these realization issues are resolved since the complicated wave propagation phenomena between metasurface layers can be accurately captured by the wave matrix approach. Generalizing our earlier theory that described only a single azimuthal order, the wave matrices, ABCD matrices, and $S$ matrices for cylindrical structures presented here account for multiple azimuthal orders, and are defined in a multimodal sense. Conversion formulas between different network parameters needed for analysis and synthesis convenience are also provided. Additionally, multimodal wave matrix expressions for the building blocks of the cascaded cylindrical metasurfaces are derived and discussed in detail. The azimuthal variation of the cylindrical metasurfaces is accounted for by a Fourier series expansion. Using this comprehensive multimodal wave matrix theory, an optimization technique is utilized to synthesize specified field transformations. In order to verify the synthesis method, the design and full-wave simulation of three interesting and powerful devices (azimuthal mode converters, illusion devices, and multi-functional metasurfaces) are conducted. These design examples demonstrate the ability to perform arbitrary field transformation using the proposed framework. Finally, several guidelines regarding practical realization are offered. Future work includes the integration of real-world feeding structures such as coaxial [\ref{Heebl_2016},\ref{Shen_1999}] or waveguide excitations, and the fabrication and measurement of prototypes.

\appendices

\section{Determination of Numerical Parameters in Synthesis Process}

In this appendix, guildlines are provided for selecting the highest azimuthal order, $M$, used in the optimization process, the number of metasurfaces, $N$, and the highest azimuthal order of variation of metasurface admittances, $K$.

To determine $M$, the large order approximation for Bessel functions is used. For a large azimuthal order $m$, the following approximation can be applied [\ref{Handbook}]:
\begin{equation}
    -jH_m^{(1)}(k\rho) \approx +jH_m^{(2)}(k\rho) \approx -\sqrt{\frac{2}{\pi m}}\Big{(}\frac{2m}{ek\rho}\Big{)}^m
.
\end{equation}

\noindent{When $m$ is sufficiently large that}
\begin{equation}
    \frac{2m}{ek\rho} \gg 1
,
\end{equation}

\noindent{Hankel functions diverge toward infinity. For most applications, the effects of such high-order modes are negligible. Taking these modes into account complicates numerical calculations without noticeable improvements in accuracy. Consequently, a reasonable choice for the highest azimuthal order $M$ is given by}
\begin{equation}
    M \approx \frac{e}{2}k\rho_\text{min}
,
\end{equation}

\noindent{where $\rho_\text{min}$ denotes the smallest radius modeled by the wave matrix theory (e.g. the radius of the innermost metasurface $\rho_1$).}

A rule of thumb is only given for the number of metasurface layers, $N$, needed since it is determined empirically. In the case of only one azimuthal order [\ref{my_TAP}], three metasurface layers are required to synthesize lossless and reciprocal scattering parameters for a single polarization. Since multiple azimuthal orders are considered in this manuscript, the problem is more complicated and naturally more sheets are needed. Therefore, in the optimization process, we start with $N=3$ layers, and gradually increase $N$ if convergence cannot be achieved.

Finally, the highest azimuthal order, $K$, of the metasurface admittance variation is also selected empirically. As discussed in Section III, when a cylindrical wave of azimuthal order $m$ interacts with a metasurface, scattered waves of orders $m-K$ and $m+K$  are generated. Since there are $N$ layers in total, $K$ is chosen to satisfy $K\times N \leq M$. This ensures that most of the higher-order modes generated by the structure are accounted for.




\begin{IEEEbiography}[{\includegraphics[width=1in,height=1.25in,clip,keepaspectratio]{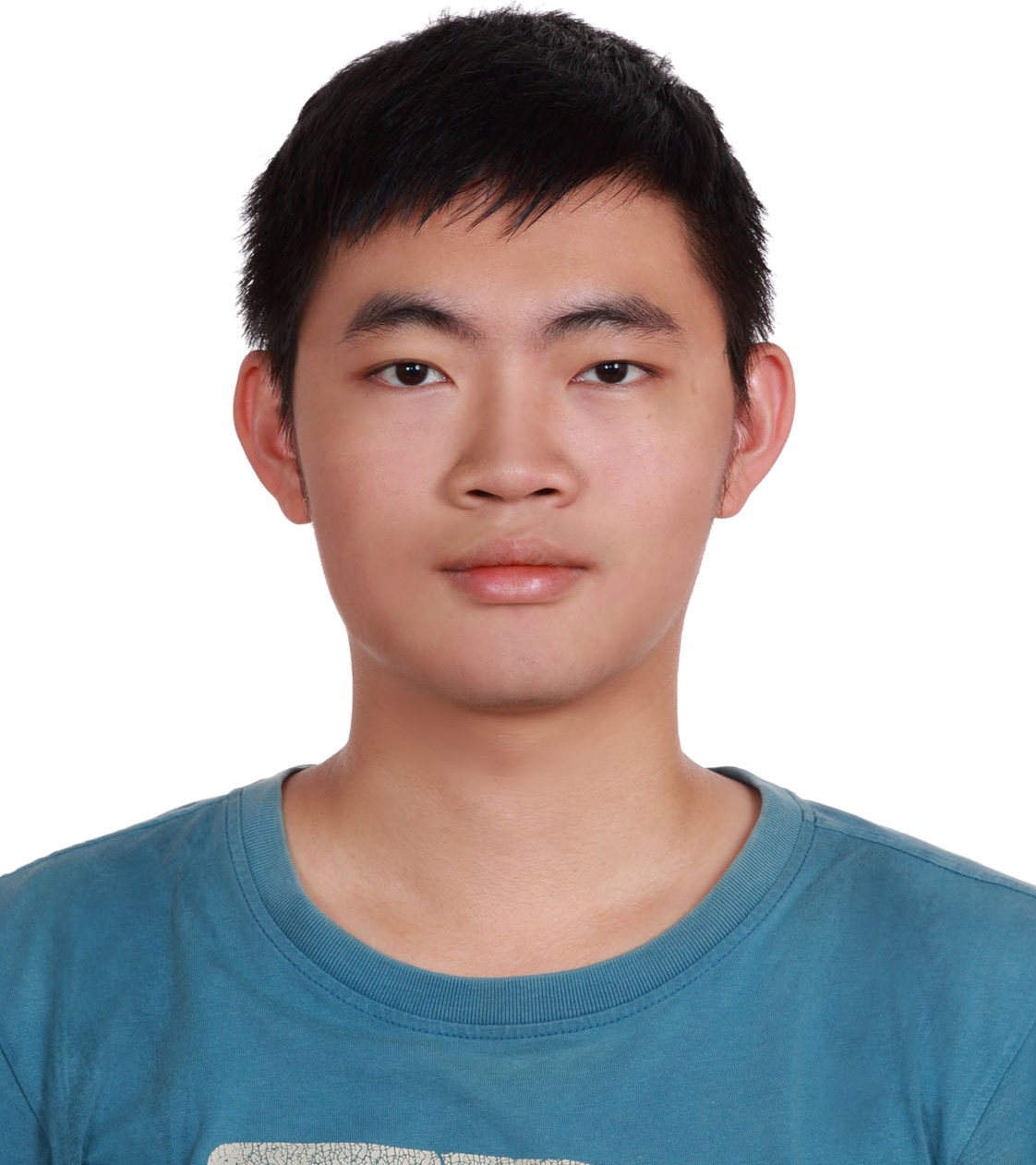}}]{Chun-Wen Lin} (S'20) received the B.S. degree in electrical engineering and M.S. degree in communication engineering from National Taiwan University, Taipei, Taiwan, in 2016 and 2018, respectively. During his study for master's degree, he was with Electromagnetic Compatibility Laboratory, National Taiwan University, as a research assistant. He was also an intern at Taiwan Semiconductor Manufacturing Company.

In 2019, Mr. Lin joined Prof. Anthony Grbic's group with the University of Michigan, Ann Arbor, MI, USA, where he is currently pursuing the Ph.D. degree in electrical and computer engineering. His research interests include metamaterials, curved metasurfaces, antenna designs, and study of analytical electromagnetics/optics. 
\end{IEEEbiography}

\begin{IEEEbiography}[{\includegraphics[width=1in,height=1.25in,clip,keepaspectratio]{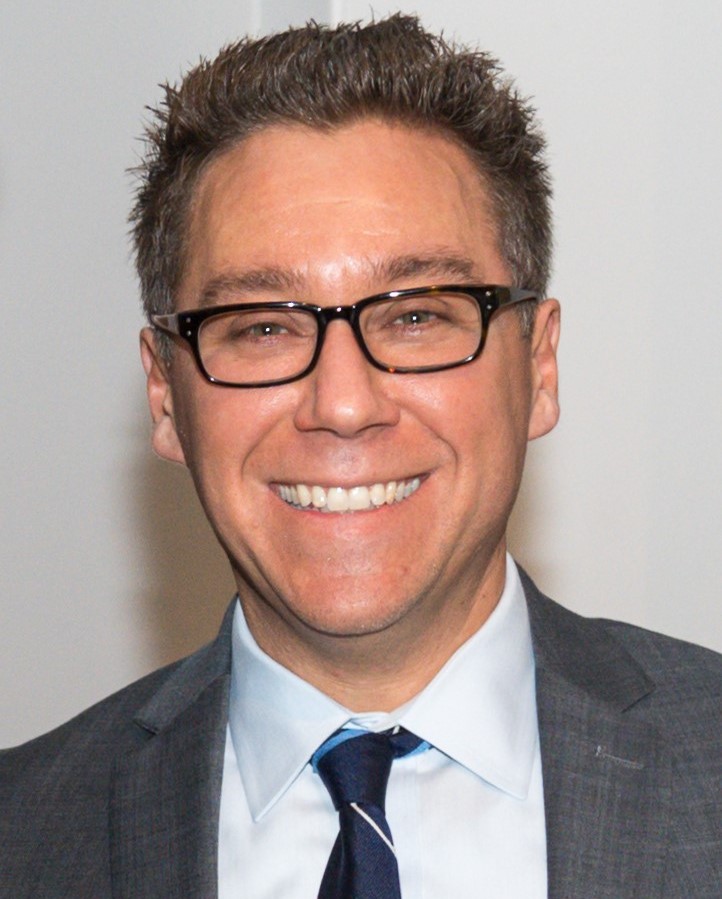}}]{Anthony Grbic} (S’00 - M’06 - SM’14 – F’16) received the B.A.Sc., M.A.Sc., and Ph.D. degrees in electrical engineering from the University of Toronto, Canada, in 1998, 2000, and 2005, respectively. In 2006, he joined the Department of Electrical Engineering and Computer Science, University of Michigan, Ann Arbor, MI, USA, where he is currently a Professor. His research interests include engineered electromagnetic structures (metamaterials, metasurfaces, electromagnetic band-gap materials, frequency-selective surfaces), antennas, microwave circuits, time varying and space-time varying electromagnetic systems, plasmonics, wireless power transmission, and analytical electromagnetics/optics.

Dr. Grbic served as Technical Program Co-Chair in 2012 and Topic Co-Chair in 2016 and 2017 for the IEEE International Symposium on Antennas and Propagation and USNC-URSI National Radio Science Meeting. He was an Associate Editor for IEEE Antennas and Wireless Propagation Letters from 2010 to 2015. Dr. Grbic was the recipient of AFOSR Young Investigator Award as well as NSF Faculty Early Career Development Award in 2008, the Presidential Early Career Award for Scientists and Engineers in January 2010. He also received an Outstanding Young Engineer Award from the IEEE Microwave Theory and Techniques Society, a Henry Russel Award from the University of Michigan, and a Booker Fellowship from the United States National Committee of the International Union of Radio Science in 2011. He was the inaugural recipient of the Ernest and Bettine Kuh Distinguished Faculty Scholar Award in the Department of Electrical and Computer Science, University of Michigan in 2012. In 2018, Prof. Grbic received a University of Michigan Faculty Recognition Award for outstanding achievement in scholarly research, excellence as a teacher, advisor and mentor, and distinguished service to the institution and profession. He is currently an IEEE Microwave Theory and Techniques Society Distinguished Microwave Lecturer and a member of the Scientific Advisory Board of the International Congress on Artificial Materials for Novel Wave Phenomena – Metamaterials.
\end{IEEEbiography}

\end{document}